\documentclass[a4paper, 12pt, reqno]{amsart}
\usepackage[applemac]{inputenc}

\usepackage{amsmath,amssymb, amsthm}
\usepackage{graphicx, color, psfrag}

\usepackage{algorithm}
\usepackage[noend]{algpseudocode}

\usepackage{breakcites}
\usepackage{url}

\usepackage{caption}
\theoremstyle{plain}
	\newtheorem{theorem}{Theorem}
	\newtheorem{proposition}[theorem]{Proposition}

\theoremstyle{definition}
	
	\newtheorem{example}[theorem]{Example}

\newcommand{\E}[1]{\mathbb{E}[#1]}
\newcommand{\Esub}[2]{\mathbb{E}_{#1}\left[#2\right]}
\newcommand{\V}[1]{\mathbb{V}[#1]}
\renewcommand{\Pr}[1]{\mathbb{P}[#1]}
\newcommand{\Pra}[1]{\mathbb{P}\left[#1\right]}
\newcommand{\Prb}[1]{\mathbb{P}\bigg[#1\bigg]}
\newcommand{\R}{\mathbb{R}}

\DeclareMathOperator{\bias}{bias}
\DeclareMathOperator{\Cov}{Cov}

\usepackage[margin=1.25in, top=1.4in, headsep=0.4in]{geometry}

\graphicspath{{./figures/}}

\begin{document}

\title{Robust Importance Sampling with Adaptive Winsorization}
\author{Paulo Orenstein}
\address{Instituto de Matem\'atica Pura e Aplicada, Estr. Dona Castorina, 110, Rio de Janeiro 22460-320, Brazil. \newline \newline \textup{\tt pauloo@impa.br}}

\begin{abstract}
Importance sampling is a widely used technique to estimate properties of a distribution. This paper investigates trading-off some bias for variance by adaptively winsorizing the importance sampling estimator. The novel winsorizing procedure, based on the Balancing Principle (or Lepskii's Method), chooses a threshold level among a pre-defined set by roughly balancing the bias and variance of the estimator when winsorized at different levels. As a consequence, it provides a principled way to perform winsorization with finite-sample optimality guarantees under minimal assumptions. In various examples, the proposed estimator is shown to have smaller mean squared error and mean absolute deviation than leading alternatives.
\end{abstract}

\maketitle

\section{Introduction} \label{sec:introduction}

 Let $\mathbb{P}$ and $\mathbb{Q}$ denote two probability measures on a set $\mathcal{X}$ with some $\sigma$-algebra, and suppose they admit probability density functions $p$ and $q$. Let $f: \mathcal{X} \to \R$ be a measurable function, integrable with respect to $\mathbb{P}$. The goal is to estimate
 \begin{equation*}
  \theta = \Esub{p}{f(X)} = \int_{\mathcal{X}} f(x) p(x) dx.
 \end{equation*}

Let $X_1,\ldots, X_n$ be an independent and identically distributed sequence of $\mathcal{X}$-valued random variables with law $q$. The importance sampling estimator for $\theta$ is defined as
\begin{equation} \label{eq:importance_sampling_estimator}
 \hat{\theta}_n = \frac{1}{n} \sum_{i=1}^{n} f(X_i) \frac{p(X_i)}{q(X_i)},
\end{equation}
with $p(x)$ being the target distribution and $q(x)$ the proposal or sampling distribution. As long as $q(x)>0$ whenever $f(x)p(x)>0$, the estimator is unbiased and, by the Weak Law of Large Numbers, consistent.

While importance sampling has many applications, from rare-event simulations to Bayesian computation, it can fail spectacularly with a poor choice of sampling distribution $q$. Indeed, because $\hat{\theta}_n$ is a ratio of random variables, it can exhibit enormous or even infinite variance.

As an example, if $\mathcal{X}=[0,\infty)$ with $f(x)=x$, $p(x)=\lambda^{-1}e^{-\lambda x}$, $\lambda>0$ and $q(x)=e^{-x}$, then the importance sampling estimator for estimating $\lambda$, the rate of the target exponential distribution, has infinite variance whenever $\lambda^{-1}\geq 2$. 
Though these estimates can be improved by developing a better suited proposal distribution, in many cases this is hard to do theoretically and can lead to distributions that are prohibitively expensive to sample from.
Section \ref{subsec:self_avoiding_walks} presents such an example in the more practical setting of estimating the number of self-avoiding walks on a grid.

A relevant question in this context is the extent to which the variance of the terms
\begin{equation*}
Y_i = f(X_i) \frac{p(X_i)}{q(X_i)}
\end{equation*}
can be controlled. While many modifications of these terms have been proposed to achieve variance reduction (see \cite{fithian2014semiparametric}, \cite{riihimaki2014laplace}, \cite{vehtari2015pareto}, \cite{delyon2016integral}, and \cite{chan2020robust}), they generally rely on delicate tail or moment conditions. This paper considers a novel way to winsorize $Y_i$ in a data-dependent manner that does not have such requirements. Indeed, define the random variables censored at levels $-M$ and $M$ by
\begin{equation*}
Y_i^M = \max(-M, \min(Y_i, M)).
\end{equation*}
With this notation, the usual importance sampling can be rewritten $\hat{\theta}_n = \frac{1}{n} \sum_{i=1}^{n} Y_i$, while the winsorized importance sampling estimator at level $M$ is
\begin{equation*}
\hat{\theta}^M_n = \frac{1}{n} \sum_{i=1}^{n} Y_i^M.
\end{equation*}

The choice of $M$ indexes a bias-variance trade-off. As the threshold level $M$ is decreased, the variance lessens as the bias grows. The extreme case $M=\infty$ chooses the sample mean, with zero bias but potentially infinite variance, while $M=0$ produces the constant estimator $\hat{\theta}_n=0$. Past proposals suggested asymptotic considerations or cross-validation as means to pick $M$ (see \cite{ionides2008truncated}, \cite{sen2017identifying}, \cite{northrop2017cross}).

To obtain finite-sample guarantees, however, this paper proposes an adaptation of the Balancing Principle (also known as Lepskii's Method, see \cite{Lepskii_1991} and \cite{mathe2006lepskii}) to pick the optimal threshold level. This method adaptively selects a level that automatically winsorizes more when the variance of the estimator is high compared to bias, and less (or not at all) when the variance is comparatively lower. The following result is a particular case of the more general Theorem \ref{thm:balanced_IS} and can be applied to the exponential example previously considered.

\begin{theorem} \label{thm:intro}
Let $\left\{Y_i\right\}_{i=1}^n$ be independent and identically distributed positive or negative random variables with mean $\theta$. Consider winsorizing $Y_i$ at different threshold levels in $\Lambda=\{M_1, \ldots, M_k\}$ to obtain winsorized samples $\{Y_i^{M_j}\}_{i=1}^n$ for $j=1, \ldots, k$. Choose the threshold level according to the rule
\begin{equation} \label{lepski_rule}
M_* = \min\left\{M \in \Lambda \ : \ \forall M', M'' \geq M, \quad  |\overline{Y^{M'}}-\overline{Y^{M''}}| \leq  ct \cdot \frac{\hat{\sigma}^{M'}+\hat{\sigma}^{M''}}{2 \sqrt{n}}\right\},
\end{equation}
with $c$, $t$ positive constants, $c>2$, $\overline{Y^M} = \frac{1}{n}\sum_{i=1}^{n}Y_i^M$ and $\hat{\sigma}^{M}=(\frac{1}{n}\sum_{i=1}^{n}(Y_i^M - \overline{Y^M})^2)^{1/2}$. 
Let $K>0$ be such that $\E{|Y_i^{M} - \E{Y_i^{M}}|^3} \leq K (\V{Y_i^{M}})^{3/2}$ for all $M \in \Lambda$ and denote by $\Phi$ the normal cumulative distribution function. Then, with probability at least
\begin{equation*}
1 - 2|\Lambda|\left(1 -\Phi\left(t \sqrt{\frac{n}{n+t^2}}\right)+ \frac{25K}{\sqrt{n}}\right),
\end{equation*}
it holds that
\begin{equation*}
|\overline{Y^{M_*}}-\theta| \leq C \min_{M \in \Lambda}\left\{|\E{Y_i^M}-\theta| +  t\cdot\frac{\hat{\sigma}^M}{\sqrt{n}}\right\},
\end{equation*}
where $C=C(c)$ can be made less than $4.25$.
\end{theorem}

Intuitively, the theorem says that, given a set $\Lambda$ of pre-chosen threshold values for winsorization, picking one according to the decision rule (\ref{lepski_rule}) guarantees with high probability that the error of the procedure is bounded by a constant multiple of the optimal sum of bias and sample standard deviation among all threshold levels considered. This oracle-type result ensures that the error is not too large in terms of either unobserved bias or sample variance. The user is free to choose parameters $c$ and $t$, and while $K$ is often unknown, it is not expected to be large in a setting where the variance of $Y_1$ can be unwieldy to begin with. Finally, the theorem requires neither tail conditions nor the existence of moments besides the mean of $Y_1$.

\begin{figure}[htpb]
	\centering
	\begin{minipage}{.48\linewidth}
	\includegraphics[width=1\textwidth]{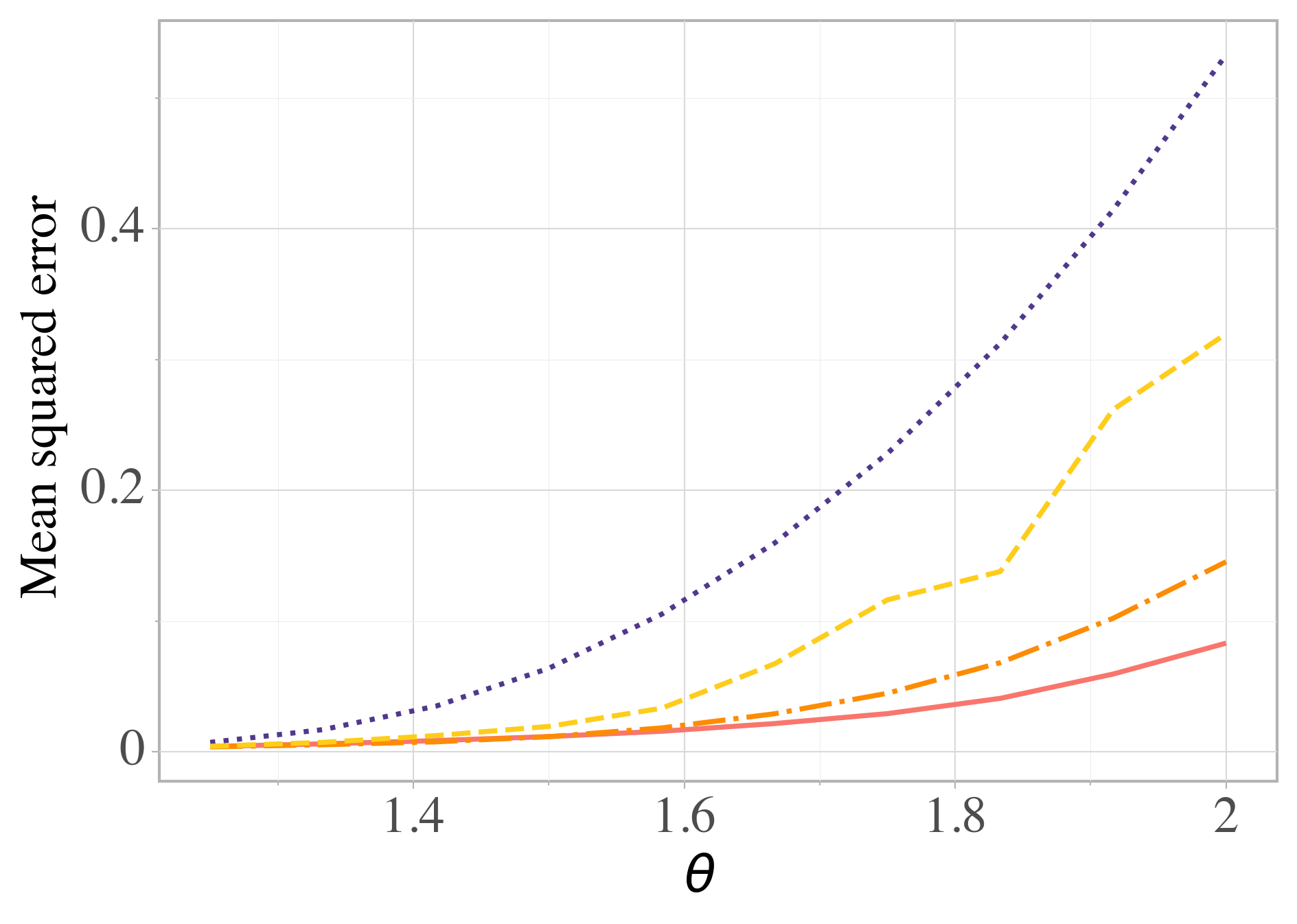}
	\end{minipage}
	\begin{minipage}{.48\linewidth}
	\includegraphics[width=1\textwidth]{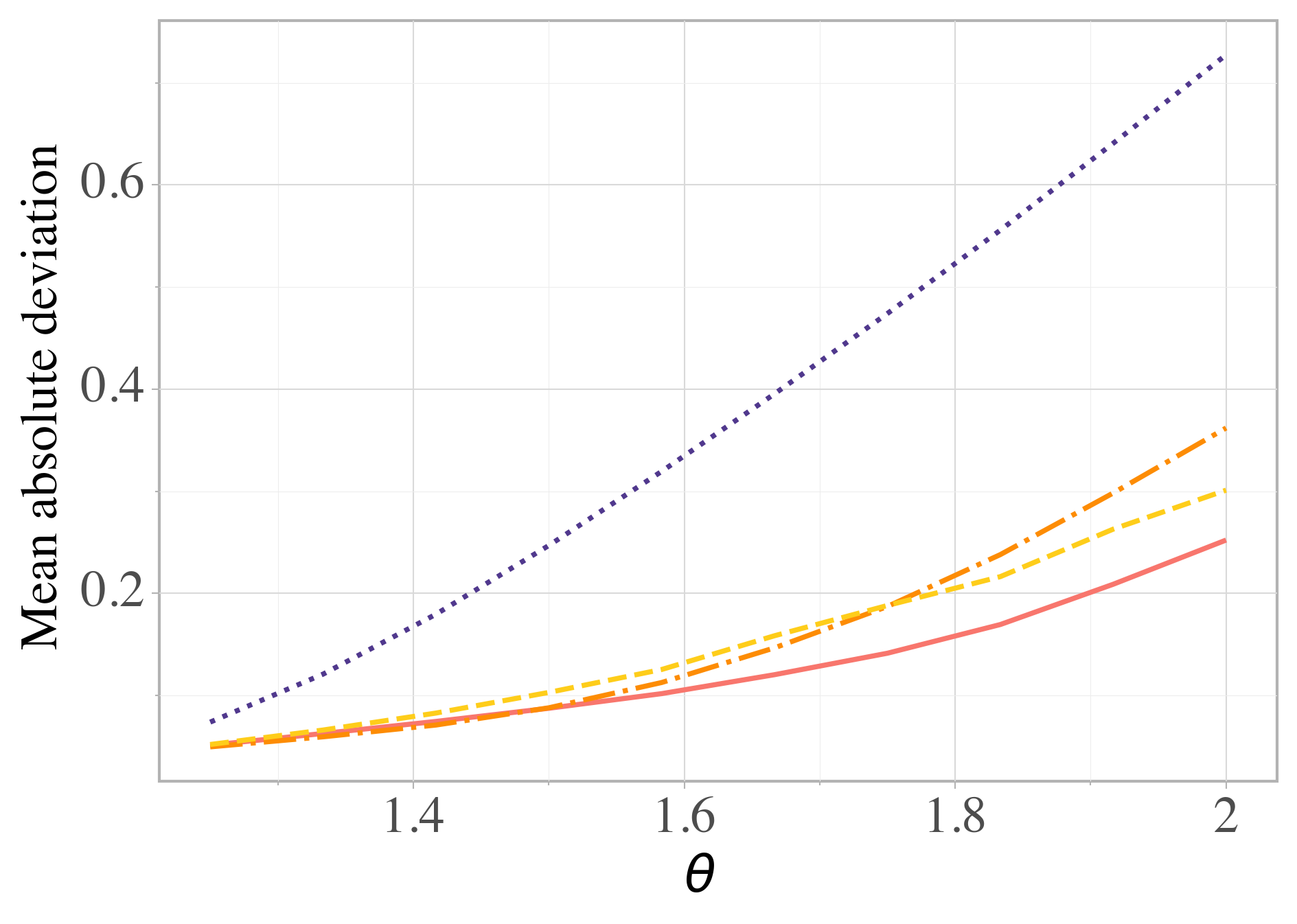}
	\end{minipage}
	\captionsetup{width=\textwidth, font=small}
	\caption{Mean squared error and mean absolute deviation errors for the exponential example, with $1.25 \leq \lambda \leq 2$, for the proposed estimator (solid line), importance sampling winsorized at a level chosen by cross-validation (dots) and at a fixed level (dot-dash), and the usual importance sampling estimator (long dash). Subsection \ref{subsec:synthetic_examples} discusses the simulation setup and includes further examples.}
	\label{fig:intro_exponential}
\end{figure}

The method can be extended to random variables that are unbounded from above and below, as discussed in Section \ref{sec:theoretical_results}, at the expense of loosening its optimality guarantees and probability bounds. Section \ref{sec:empirical_results} shows that this theoretical guarantee translates into good performance both in terms of mean squared error as well as mean absolute deviation. For instance, Figure \ref{fig:intro_exponential} displays the mean squared error and mean absolute errors when winsorizing the importance sampling estimator at level (\ref{lepski_rule}) in the exponential example with increasing values of $\lambda$, against other importance sampling estimators. Subsection \ref{subsec:synthetic_examples} includes the simulation details.

\section{Theoretical Results} \label{sec:theoretical_results}

At its heart, the proposed method relies on increasing the threshold level $M$ to decrease variance while adding some bias. It is always the case that further winsorizing never increases variance.

\begin{proposition} \label{prop:variance}
Let $Y_1, \ldots, Y_n$ be independent and identically distributed random variables with finite mean. Denote the $M$-winsorized version of each $Y_i$ by $Y_i^M = \max(-M, \min(Y_i, M))$, for $i=1, \ldots, n$ and $M\geq0$. If $\hat{\theta}_n = \frac{1}{n}\sum_{i=1}^{n} Y_i$ and $\hat{\theta}^M_n = \frac{1}{n} \sum_{i=1}^{n} Y_i^M$, then
\begin{equation*}
\V{\hat{\theta}^M_n} \leq \V{\hat{\theta}_n}.
\end{equation*}
Furthermore, $\V{\hat{\theta}^{M}_n}$ is a non-decreasing function of $M$.
\end{proposition}

\begin{proof}
The inequality trivially holds if $Y_1$ has infinite variance, so assume it is finite. Note $\V{\hat{\theta}_n} = \frac{1}{n}\V{Y_1}$ and $\V{\hat{\theta}^M_n} = \frac{1}{n}\V{Y_1^M}$, so it remains to show that $\V{Y_1^M} \leq \V{Y_1}$. Since
\begin{equation*}
	\V{Y_1} + \V{Y^M_1} - 2 \Cov(Y_1, Y_1^M) = \V{Y_1-Y_1^M} \geq 0,
\end{equation*}
the results follows if $\Cov(Y_1, Y_1^M) \geq \V{Y_1^M}$. By definition of $M$-winsorization, $Y_1^M (Y_1 - Y_1^M) = M|Y_1 - Y_1^M|$, so
\begin{align*}
\Cov(Y_1, Y_1^M) -\V{Y_1^M}
	&= \E{Y_1^M(Y_1-Y_1^M)} - \E{Y_1^M}\E{Y_1-Y_1^M} \\
	&= \E{M|Y_1-Y_1^M|} - \E{Y_1^M}\E{Y_1-Y_1^M}\\
	&\geq \E{M|Y_1-Y_1^M|}-\E{|Y_1^M|}\E{|Y_1-Y_1^M|} \\
	&\geq 0,
\end{align*}
where the last inequality follows from $\E{|Y_1^M|}\leq M$ since $|Y_1^{M}| \leq M$.

To show $\V{\hat{\theta}_n^M}$ is  non-decreasing in $M$, take two winsorization levels $M''\geq M' \geq 0$, and consider the variance of the sample $\tilde{Y}_i=Y^{M''}_i$, as well as its $M'$-winsorization $\tilde{Y}^{M'}_i=Y^{M'}_i$. The above implies $\V{Y^{M''}_1}=\V{\tilde{Y}_1} \geq \V{\tilde{Y}^{M'}_1}=\V{Y^{M'}_1}$, so $\V{Y^M_1}$ is non-decreasing with $M$.

These results hold in far greater generality. In particular, winsorization decreases every centered absolute moment of order greater than one, even for non-symmetric forms of winsorization. See \cite{chow1969} for a general proof based on convexity.
\end{proof}

On the other hand, the bias is not always monotonically increasing with $M$ (the Supplementary Material includes a pathological but simple counter-example). However, the following proposition establishes the conditions under which this is true.

\begin{proposition} \label{prop:bias}
Let $Y_1, \ldots, Y_n$ be independent and identically distributed, with mean $\theta \in \R$, and let $Y_i^M = \max(-M, \min(M, Y_i))$ be its $M$-winsorization. Then, 
\begin{equation*}
	\bias(M) = |\E{Y_1^M}-\theta| = \left|\int_{M}^{\infty} ( \Pr{Y > t} - \Pr{Y \leq -t}) dt \right|.
\end{equation*}
A sufficient condition for bias to be non-increasing is that $Y$ stochastically dominates $-Y$ or vice-versa, which is implied, in particular, if $Y$ is positive, negative or symmetric.
\end{proposition}

\begin{proof}
Since, for $t>0$, $\Pr{Y_1^M > t} = \Pr{Y_1 > t}$ when $t < M$ and zero otherwise,
\begin{align*}
	\E{Y_1^M} &= \int_{0}^{\infty} \Pr{Y_1^M > t} dt - \int_{-\infty}^{0} \Pr{Y_1^M \leq t} dt \\
	&= \int_{0}^{M} \Pr{Y_1>t} dt - \int_{-M}^{0} \Pr{Y_1 \leq t} dt,
\end{align*}
which yields the following expression for the bias:
\begin{align*}
	|\E{Y_1^M}-\E{Y_1}| &= \left| \int_{0}^{\infty} (\Pr{Y_1^M > t} - \Pr{Y_1 > t}) dt - \int_{-\infty}^{0} (\Pr{Y_1^M \leq t}-\Pr{Y_1 \leq t}) dt\right| \\
	&= \left|\int_{M}^{\infty} \Pr{Y>t} dt - \int_{-\infty}^{-M} \Pr{Y\leq t} dt\right| \\
	&= \left|\int_{M}^{\infty} (\Pr{Y>t}- \Pr{Y\leq -t}) dt  \right|.
\end{align*}

Thus, a sufficient condition for the bias to be non-increasing is $\Pr{Y \geq t} \geq \Pr{Y \leq -t}$ for all $t \geq 0$ or $\Pr{Y \leq -t} \geq \Pr{Y\geq t}$ for all $t \geq 0$, that is, if $Y$ stochastically dominates $-Y$ or vice-versa. This trivially holds when $Y$ is positive, negative or symmetric.
\end{proof}

Assuming the bias is non-increasing and the variance is non-decreasing, how to select an appropriate level $M$? If the actual bias and variance incurred by the winsorized importance sampling estimator, $|\E{\hat{\theta}^M} - \theta|$ and $\V{\hat{\theta}^M}$, were known, one could pick the thresholding level to exactly balance out the increasing effects winsorizing has on the bias and the decreasing effects it has on the variance. More concretely, one could consider using the sample versions of these quantities: to choose between two thresholding levels $M'>M''$, if the increase in sample bias $|(\frac{1}{n}\sum_{i=1}^{n}Y_i^{M'}- \theta)-(\frac{1}{n}\sum_{i=1}^{n}Y_i^{M''}-\theta)|$ is small enough relative to the standard error, then it might be worth winsorizing the sample at the lower level $M''$. This idea can be developed into a novel adaptation of the Balancing Principle, which is useful in its own right (see the Supplementary Material for a proof and \cite{hamarik2009balancing} for an overview of the method in more abstract settings).

\begin{proposition}[Balancing Principle, adapted] \label{prop:balancing}
Suppose $\theta \in \R$ is an unknown parameter, $\{\hat{E}^M\}_{M \in \Lambda}$ is a sequence of estimators of $\theta$ indexed by $M \in \Lambda \subset \R$, with $\Lambda$ a non-empty finite set. Additionally, suppose that $|\hat{E}^M - \theta| \leq b(M)+\hat{s}(M)$ for each $M \in \Lambda$, where $b(M) \geq 0$ is an unknown but non-increasing function in $M$, while $\hat{s}(M) \geq 0$ is observed and non-decreasing in $M$. Choose $c>2$, and take
\begin{equation} \label{eq:balancing_rule}
 M_* = \min\left\{M \in \Lambda : \forall M', M'' \geq M, \ |\hat{E}^{M'}-\hat{E}^{M''}| \leq c
 \left(\frac{\hat{s}(M')+\hat{s}(M'')}{2}\right)\right\}.
\end{equation}
Then,
\begin{equation} \label{eq:balancing_guarantee}
 |\hat{E}^{M_*}-\theta| \leq C \min_{M \in \Lambda} \left\{b(M)+\hat{s}(M)\right\},
\end{equation}
where $C$ only depends on the chosen value for $c$.
\end{proposition}

\begin{proof}
Since $\max\left\{M: M \in \Lambda\right\}$ is always a candidate for $M_*$, the minimum is well defined. The goal is to show there exists $C \geq 0$ such that, for all $M \in \Lambda$, $|\hat{E}^{M_*}-\theta| \leq C(\hat{s}(M)+ b(M))$. For this consider two cases.
 
\vspace{5mm}
 
(i) First, let $M$ be such that $M\geq M_*$. By definition of $M_*$, and since $\hat{s}(M)$ is non-decreasing in $M$,
\begin{equation*}
 |\hat{E}^{M_*}-\hat{E}^{M}| \leq c \cdot \left(\frac{\hat{s}(M)+\hat{s}(M_*)}{2}\right) \leq c\hat{s}(M).
\end{equation*}
By assumption $|\hat{E}^{M}-\theta|\leq b(M)+ \hat{s}(M)$, so, from the triangle inequality,
\begin{equation*}
|\hat{E}^{M_*}-\theta| \leq |\hat{E}^{M_*}-\hat{E}^{M}| + |\hat{E}^{M}-\theta| \leq c\hat{s}(M) + b(M) + \hat{s}(M) \leq b(M)+\left(1+c\right)\hat{s}(M).
\end{equation*}
This proves the case $M\geq M_*$ with $C=1+c$.

\vspace{5mm}

(ii) Now suppose $M<M_*$. In that case, since $\Lambda$ is finite, consider the sequence of elements in $\Lambda$ in between, namely $M=M_i<M_{i+1}<\cdots<M_{j-1}<M_j=M_*$. Then, there exist $M', M'' \geq M_{j-1}$ such that $|\hat{E}^{M'}-\hat{E}^{M''}| > c (\hat{s}(M')+\hat{s}(M''))/2$, where either $M'=M_{j-1}$ or $M''=M_{j-1}$ (otherwise contradicting the minimality of $M_*$). Without loss of generality, assume $M''=M_{j-1}$. This then implies $M' \geq M_*$, since $M'=M_{j-1}$ gives another contradiction: $0=|\hat{E}^{M'}-\hat{E}^{M''}|>c(\hat{s}(M')+\hat{s}(M''))/2>0$. Hence,
\begin{equation} \label{eq7}
|\hat{E}^{M'}-\hat{E}^{M''}| > c \left(\frac{\hat{s}(M')+\hat{s}(M'')}{2}\right).
\end{equation}
Also, by the triangle inequality,
\begin{equation} \label{eq8}
|\hat{E}^{M''} - \hat{E}^{M'}| \leq |\hat{E}^{M''} - \theta| + |\hat{E}^{M'}-\theta| \leq \left(b(M'')+b(M')\right) + \left(\hat{s}(M'')+\hat{s}(M')\right),
\end{equation}
so, combining (\ref{eq7}) and (\ref{eq8}) yields
\begin{align}
c \left(\frac{\hat{s}(M')+\hat{s}(M'')}{2}\right) &< |\hat{E}^{M''}-\hat{E}^{M'}| \leq \left(b(M'')+b(M')\right) + \left(\hat{s}(M'')+\hat{s}(M')\right). \label{eq9}
\end{align}
Recall that $M' \geq M_* > M'' \geq M$, so 
\begin{align} \label{eq10}
 b(M') \leq b(M_*) &\leq b(M'') \leq b(M) \\
 \hat{s}(M') \geq \hat{s}(M_*) &\geq \hat{s}(M'') \geq \hat{s}(M).
\end{align}
From (\ref{eq9}) and (\ref{eq10}),
\begin{equation*}
\left(\frac{c}{2}-1\right)(\hat{s}(M')+\hat{s}(M''))  \leq b(M'') + b(M')
\leq 2 \ b(M),
\end{equation*}
and, because $c>2$ and $\hat{s}(M'') \geq 0$,
\begin{equation} \label{eq:bound_on_s_by_b}
\hat{s}(M') \leq \frac{4}{c-2} b(M).
\end{equation}
Finally,
\begin{equation*}
|\hat{E}^{M_*}-\theta| \leq b(M_*) + \hat{s}(M_*) \stackrel{(\ref{eq10})}{\leq} b(M) + \hat{s}(M') \stackrel{(\ref{eq:bound_on_s_by_b})}{\leq} \left(1+\frac{4}{c-2}\right)b(M),
\end{equation*}
which naturally implies
\begin{equation*}
 |\hat{E}^{M_*} - \theta| \leq \left(1 + \frac{4}{c-2}\right)\left(b(M) + \hat{s}(M)\right),
\end{equation*}
and the second case is proved with $C=1+4/(c-2)$. Thus, since $C(c) \geq \max_{c>2} \left\{c+1, 1+4/(c-2)\right\}$, $C$ is minimized at $c=1+\sqrt{5}$, where $C=2+\sqrt{5} < 4.25$.
\end{proof}

For the importance sampling setting, consider the following choices.
\begin{align*}
	\hat{E}^{M} &= \overline{Y^{M}} = \frac{1}{n} \sum_{i=1}^{n} Y_i^{M}, \\
	\hat{s}(M) &= t \cdot \frac{\hat{\sigma}^{M} }{\sqrt{n}} = t \cdot \frac{1}{n} \sqrt{\sum_{i=1}^{n} (Y_i^{M}-\overline{Y^{M}})^{2}}, 
\end{align*}
where $t>0$ is a chosen constant. To apply the theorem above, it is necessary to have $|\overline{Y^{M}} - \theta| \leq b(M)+t \hat{\sigma}^{M}/\sqrt{n}$ for all $M \in \Lambda$, for some non-increasing function $b(M)$. Naturally this can only hold probabilistically in the importance sampling setting, and if $b(M)$ is some upper bound on the bias $|\E{Y_1^{M}} - \theta|$ then these probabilities can be calculated via the limit theory for self-normalized sums, yielding the following result.

\begin{theorem} \label{thm:balanced_IS}
Let $\left\{Y_i\right\}_{i=1}^n$ be independent random variables with mean $\theta$. Consider winsorizing $Y_i$ at different threshold levels in a pre-chosen set $\Lambda=\{M_1, \ldots, M_k\}$ to obtain winsorized samples $\{Y_i^{M_j}\}_{i=1}^n$, $j=1, \ldots, k$. Pick the threshold level according to the rule
\begin{equation} \label{eq:balanced_IS_rule}
M_* = \min\left\{M \in \Lambda \ : \ \forall M', M'' \geq M, \quad  |\overline{Y^{M'}}-\overline{Y^{M''}}| \leq ct \cdot \frac{\hat{\sigma}^{M'}+\hat{\sigma}^{M''}}{2 \sqrt{n}}\right\},
\end{equation}
where $c, t$ are positive constants with $c>2$. Let $K>0$ be such that $\E{|Y_i^{M} - \E{Y_i^{M}}|^3} \leq K (\V{Y_i^{M}})^{3/2}$ for all $M\in \Lambda$ and denote by $\Phi$ the normal cumulative distribution function. If $b(M)$ is any non-increasing upper bound on the bias $|\E{Y_1^M}-\theta|$, with probability at least
\begin{equation} \label{eq:balanced_IS_prob}
1 - 2|\Lambda|\left(1 -\Phi\left(t \sqrt{\frac{n}{n+t^2}}\right)+ \frac{25K}{\sqrt{n}}\right),
\end{equation}
it holds that
\begin{equation} \label{eq:balanced_IS_guarantee_general}
	|\overline{Y^{M_*}}-\theta| \leq C \min_{M \in \Lambda}\left\{b(M) +  t \cdot \frac{\hat{\sigma}^M}{\sqrt{n}}\right\},
\end{equation}
where $C=C(c)$ is less than $4.25$ when $c=1+\sqrt{5}$.

In particular, if $Y_1$ stochastically dominates $-Y_1$ or vice-versa, then is possible to take $b(M)=|\E{Y_1^M}-\theta|$ in (\ref{eq:balanced_IS_guarantee_general}), yielding
\begin{equation} \label{eq:balanced_IS_guarantee}
|\overline{Y^{M_*}}-\theta| \leq C \min_{M \in \Lambda}\left\{|\E{Y_i^M}-\theta| +  t \cdot \frac{\hat{\sigma}^M}{\sqrt{n}}\right\}.
\end{equation}
\end{theorem}

\begin{proof}
The result in (\ref{eq:balanced_IS_guarantee_general}) follows from Proposition \ref{thm:balanced_IS} with
\begin{align*}
	\hat{E}^{M} &= \overline{Y^{M}} = \frac{1}{n} \sum_{i=1}^{n} Y_i^{M}, \\
	\hat{s}(M) &= \frac{t}{n^{1/2}} \hat{\sigma}^{M} = \frac{t}{n^{1/2}} \cdot  \left(\frac{1}{n}\sum_{i=1}^{n} (Y_i^{M}-\overline{Y^{M}})^{2}\right)^{1/2},
\end{align*}
as long as $|\overline{Y^M} - \theta| \leq b(M) +t \hat{\sigma}^{M}/n^{1/2}$ for all $M \in \Lambda$, for some non-increasing function $b(M)$. By hypothesis, $b(M) \geq |\E{Y_1^{M}}-\theta|$,  so it suffices to investigate the bound when $b(M)=|\E{Y_1^M}-\theta|$. In this case, if $Y_1$ stochastically dominates $-Y_1$ or vice-versa, then Proposition \ref{prop:bias} guarantees $|\E{Y_1^M}-\theta|$ is non-increasing and therefore a valid choice for $b(M)$, implying (\ref{eq:balanced_IS_guarantee}) from (\ref{eq:balanced_IS_guarantee_general}).

Hence, it is sufficient to show that (\ref{eq:balanced_IS_prob}) bounds the probabilitty that $|\overline{Y^{M}}-\theta|\leq |\E{Y_{1}^{M}}-\E{Y_{1}}| + t \hat{\sigma}^{M}/n^{1/2}$ for all $M \in \Lambda$. To do this, first consider bounding the probability for a single $M$. By the reverse triangle inequality,
\begin{align} \label{eq:a13}
	\Prb{&\left|\overline{Y^M}-\E{Y_1} \right| \leq \left|\E{Y_1^M} - \E{Y_1}\right| + \frac{t}{n^{1/2} }\hat{\sigma}^M} \\
	&\geq \Pra{\left|\frac{1}{n}\sum_{i=1}^{n} Y_i^M - \E{Y_1^M}\right| \leq \frac{t}{n^{1/2}}\hat{\sigma}^M}.
\end{align}
Define the demeaned random variable
\begin{equation*}
	Z_i = Y_i^M - \E{Y_1^M},
\end{equation*}
so that
\begin{align} \label{eq:a14}
	\Pra{\left|\frac{1}{n}\sum_{i=1}^{n} Y_i^M - \E{Y_1^M} \right| \leq \frac{t}{n^{1/2}}\hat{\sigma}^{M}}
	&= \Pra{ \frac{n^{1/2}|\overline{Z}|}{\left(\frac{1}{n}\sum_{i=1}^{n} (Z_i-\overline{Z})^2\right)^{1/2} } \leq t}.
\end{align}
Let $h(z) = n^{1/2} z /(n+z^{2})^{1/2}$, and note that $h(z)$ is an increasing function and
\begin{align*}
	h\left(\frac{n^{1/2} |\overline{Z}| }{\left(\frac{1}{n}\sum_{i=1}^{n} (Z_i - \overline{Z})^{2}\right)^{1/2}}\right)
	&= \frac{n |\overline{Z}|}{\left(\sum_{i=1}^{n} (Z_i - \overline{Z})^{2} + n |\overline{Z}|^{2}\right)^{1/2}} 
	= \left|\frac{\sum_{i=1}^{n} Z_i}{\left(\sum_{i=1}^{n} Z_i^{2}\right)^{1/2}}\right|,
\end{align*}
so, applying $h$ to both sides, 
\begin{align*}
	\Pra{\frac{n^{1/2} |\overline{Z}|}{\left(\frac{1}{n} \sum_{i=1}^{n} (Z_i - \overline{Z})^{2}\right)^{1/2}} \leq t} &= \Pra{ \left|\frac{\sum_{i=1}^{n} Z_i}{\left(\sum_{i=1}^{n} Z_i^{2}\right)^{1/2}}\right| \leq \frac{n^{1/2} t}{\left(n+t^{2}\right)^{1/2} }} \\
	&= \Pra{\frac{\sum_{i=1}^{n} Z_i}{\left(\sum_{i=1}^{n} Z_i^{2}\right)^{1/2}} \leq \frac{n^{1/2} t}{\left(n + t^{2}\right)^{1/2}}} \\
	& \qquad - \Pra{\frac{\sum_{i=1}^{n} Z_i}{\left(\sum_{i=1}^{n} Z_i^{2}\right)^{1/2}}\leq - \frac{n^{1/2} t}{\left(n+t^{2}\right)^{1/2}}}.
\end{align*}

To bound each of the two terms in the right hand side, the following self-normalized inequality of \cite[Theorem 1.1]{shao2005explicit}, reproduced below, is useful.

\begin{theorem}[\cite{shao2005explicit}] Let $Z_1, \ldots, Z_n$ be a sequence of independent random variables with $\E{Z_i}=0$ and $\E{Z_i^2}<\infty$. Then
	\begin{equation*}
		\sup_z \bigg| \Pra{\frac{\sum_{i=1}^{n}Z_i}{\left(\sum_{i=1}^{n}Z_i^2\right)^{1/2}}\leq z} - \Phi(z)\bigg| \leq 25 \frac{\sum_{i=1}^{n} \E{|Z_i|^3}}{\left(\sum_{i=1}^{n} \E{Z_i^2}\right)^{3/2}}.
	\end{equation*}
\end{theorem}

If $K>0$ is such that $\E{|Z_i|^{3}}\leq K (\E{Z_i^{2}})^{3/2}$, then
\begin{equation*}
	25 \frac{\sum_{i=1}^{n} \E{|Z_i|^{3}}}{\left(\sum_{i=1}^{n} \E{Z_i^{2}}\right)^{3/2}}  \leq 25 \frac{Kn (\E{Z_1^{2}})^{3/2}}{n^{3/2}(\E{Z_1^{2}})^{3/2}} = \frac{25 K}{n^{1/2}},
\end{equation*}
yielding the following bounds:
\begin{align*}
	\Pra{\frac{\sum_{i=1}^{n} Z_i}{\left(\sum_{i=1}^{n} Z_i^{2}\right)^{1/2}} \leq \frac{n^{1/2}t}{\left(n+t^{2})^{1/2}\right)^{1/2}}} &\geq \Phi\left(\frac{n^{1/2}t}{\left(n+t^{2}\right)^{1/2}}\right) - \frac{25K}{n^{1/2}}, \\
	\Pra{\frac{\sum_{i=1}^{n} Z_i}{\left(\sum_{i=1}^{n} Z_i^{2}\right)^{1/2}} \leq - \frac{n^{1/2}t}{\left(n+t^{2}\right)^{1/2}}} &\leq
	1 - \Phi\left(\frac{n^{1/2}t}{\left(n+t^{2}\right)^{1/2}}\right) + \frac{25K}{n^{1/2}}.
\end{align*}
Hence,
\begin{equation} \label{eq:a15}
	\Pra{\frac{n^{1/2}|\overline{Z}|}{\left(\frac{1}{n}\sum_{i=1}^{n} (Z_i-\overline{Z})^{2}\right)^{1/2}} \leq t} \geq 
	1 - 2 \left(1 - \Phi\left(\frac{n^{1/2}t}{\left(n+t^{2}\right)}\right)+\frac{25K}{n^{1/2}}\right),
\end{equation}
and putting (\ref{eq:a13}), (\ref{eq:a14}) and (\ref{eq:a15}) together,
\begin{align*}
	\Prb{&|\overline{Y^M} - \E{Y_1}| \leq |\E{Y^M_1} - \E{Y_{1}}| + t \hat{\sigma}^M/n^{1/2}} \\
	&\geq 1 - 2\left(1 - \Phi\left(\frac{n^{1/2}t}{\left(n+t^{2}\right)^{1/2}}\right) + \frac{25K}{n^{1/2}}\right).
\end{align*}
Since this condition must hold for all $M \in \Lambda$, by the union bound,
\begin{align*}
	\Prb{&\left|\overline{Y^{M}}-\E{Y_{1}}\right| \leq \left|\E{Y_{1}^{M}-\E{Y_{1}}}\right| + \frac{t}{n^{1/2}}\hat{\sigma}^{M}, \quad \forall M \in \Lambda} \\
	&\geq 1 - \sum_{M \in \Lambda} \Pra{\left|\overline{Y^M}-\E{Y_1}\right| > \left|\E{Y_1^M}-\E{Y_1}\right| + \frac{t}{n^{1/2}}\hat{\sigma}^M}\\
	&=  1 - \sum_{M \in \Lambda} \left(1 - \Pra{\left|\overline{Y^M} - \E{Y_1}\right| \leq \left|\E{Y_1^M}-\E{Y_1}\right| + \frac{t}{n^{1/2}}\hat{\sigma}^M}\right) \\
	&\geq  1 - 2 |\Lambda| \left(1 - \Phi\left(\frac{n^{1/2}t}{\left(n+t^{2}\right)^{1/2}}\right) + \frac{25K}{n^{1/2}}\right),
\end{align*}
proving the theorem.
\end{proof}

The theorem above gives an effective, principled way to perform winsorization with finite-sample guarantees akin to oracle inequalities. While it requires a large sample size for the probability (\ref{eq:balanced_IS_prob}) to be substantial (for example, if $|\Lambda|=5$, $K=4$ and $t=3$, then $n=10^{8}$ is needed for the probability to be about $90\%$), this bound is generally very loose, as should be clear from the proof, which is also the case for bound (\ref{eq:balanced_IS_guarantee_general}).
Furthermore, there are no tail or moment conditions on the distribution of $Y_1$ besides the existence of a mean, and the result (\ref{eq:balanced_IS_guarantee}) also holds if $Y_1$ can be upper or lower bounded by any constant, not just zero, by a translation argument. While the value of $K$ is hard to know in practice, its effect can be offset by larger values of $n$ and should be small in high variance settings. Finally, the theorem suggests an algorithmic procedure less computationally intensive than similar estimating techniques while retaining competitive empirical performance, as shown in Section \ref{sec:empirical_results}.

\section{Algorithm and Practical Considerations} \label{sec:algorithm}

The theoretical results in the previous section immediately lend themselves to a practical implementation. Algorithm \ref{alg:balancing_algorithm} receives data $\{y_i\}_{i=1}^{n}$, threshold set $\Lambda$ and a constant $c$ with default value of $1+\sqrt{5}$, and output the optimal threshold $M_*$ to be used in winsorizing the data.

\begin{algorithm}
\caption{Balancing Algorithm}\label{alg:balancing_algorithm}
\begin{algorithmic}[1]
\Procedure{balancing\_algorithm}{$\left\{y_i\right\}_{i=1}^n$, $\Lambda$, $c=1+\sqrt{5}$}
\State Sort elements in $\Lambda$ to obtain $M_{(1)} \leq M_{(2)} \leq \cdots \leq M_{(k)}$
\State Initialize $k \leftarrow |\Lambda|$, $M_* \leftarrow M_{(k)}$, $i \leftarrow k-1$
\While{$|\hat{E}^{M_{(i)}}-\hat{E}^{M_{(j)}}| \leq c(\hat{s}(M_{(i)})+\hat{s}(M_{(j)}))/2$, for $j = i+1, \ldots, k$} \label{alg:while_line}
\If{$i=1$}
\textbf{Return} $M_{(i)}$
\EndIf
\State $M_* \leftarrow M_{(i)}$
\State $i \leftarrow i-1$
\EndWhile
\State \textbf{Return} $M_*$
\EndProcedure
\end{algorithmic}
\end{algorithm}

With $n$ datapoints and $k=|\Lambda|$ threshold values, the number of operations is bounded above by $O(k(k+n))$. While the algorithm is of quadratic order, in general not all values in $\Lambda$ have to be considered, speeding the method considerably.

The set of candidate threshold values $\Lambda$ is important. While the size of $\Lambda$ does not impact the guarantee (\ref{eq:balanced_IS_guarantee}), it affects the probabilistic bound (\ref{eq:balanced_IS_prob}) and the algorithm running time. Furthermore, including only extremely high or low threshold values can nudge the procedure towards over- or under-winsorizing. On the other hand, this specification is not unlike other popular procedures such as cross-validation, and allows the user to encode prior information in the choice of threshold levels.
A conservative proposal is to pick a sequence of exponentially decreasing weights starting from a high value that encourages little to no winsorizing (see \cite{mathe2006lepskii}) or to select scalar multiples of $\sqrt{n}$, following the asymptotic considerations in \cite{ionides2008truncated}. Section \ref{sec:empirical_results} employs both of these approaches in applications and finds that they give good results.

The algorithm can be extended in several directions. First, note the functional $\Phi(\hat{s}(M'), \hat{s}(M''))=(\hat{s}(M')+\hat{s}(M''))/2$ in (\ref{eq:balancing_rule}) can take other forms without affecting the proof of Theorem \ref{prop:balancing}; for example, $\Phi(\hat{s}(M'), \hat{s}(M''))=\max(\hat{s}(M'), \hat{s}(M''))$ results in a better constant $C=2+\sqrt{3}$ but in practice over-winsorizes relative to the sample mean. The choice of functions $b(M)$ and $\hat{s}(M)$ are arbitrary as long as they retain their monotonicity. Finally, the algorithm does not need to winsorize around zero (but should ideally winsorize around the mean, which is assumed to be hard to estimate in the first place).

\section{Empirical Results} \label{sec:empirical_results}

\subsection{Synthetic Examples} \label{subsec:synthetic_examples}

To study the performance of the proposed method, consider four different importance sampling examples where a sample of $n$ data points from proposal density $q$ is used to estimate the mean of a function $f$ under the target density $p$:
\begin{itemize}
	\item Beta: $f(x)=(x \cdot (1-x))^{-1/2}$, with $p$ the density of $\text{Unif}[0,1]$ and $q$ that of $\text{Beta}(\theta, \theta)$, $\theta \in [0.8, 1]$;
	\item Chi squared: $f(x)=1$, with $p$ the density of $\chi^{2}_{50}$ and $q$ that of $\chi^{2}_{\theta}$, $\theta \in [66, 75]$;
	\item Exponential: $f(x)=x$, with $p$ the density of $\text{Exp}(1/\theta)$ and $q$ that of $\text{Exp}(1)$, $\theta \in [1.25, 2]$;
	\item Normal: $f(x)=x^{2}$, with $p$ the density of $N(0,1)$ and $q$ that of $N(0, (1+\theta)^{-1})$, $\theta \in [0.6, 1]$.
\end{itemize}
In each of these examples, a parameter $\theta$ indexes how hard the problem is, in the sense that growing $\theta$ implies higher variance for the usual importance sampling estimator.

The following estimators are compared: (i) balanced importance sampling, the winsorized estimator proposed in this paper via the choice (\ref{eq:balanced_IS_rule}); (ii) cross-validated importance sampling, a similar estimator but with the winsorization level chosen via cross-validation; (iii) $\sqrt{n}$-winsorized importance sampling, where the winsorization level is kept at $\sqrt{n}$, motivated by the suggestion in \cite{ionides2008truncated}; and (iv) the usual importance sampling estimator.

For each example and choice of $\theta$, $n=1,000$ samples are generated from the proposal distribution and the importance sampling estimates are calculated. The mean squared error and mean absolute deviation of these estimates are obtained by averaging the results of $10,000$ repetitions. For the proposed balanced importance sampling, $c=1+\sqrt{5}$ is used and $t=1/\sqrt{n}$, making the procedure less willing to winsorize as $n$ grows (this gives a poor probabilistic upper bound (\ref{eq:balanced_IS_prob}) but a much better guarantee (\ref{eq:balanced_IS_guarantee}); recall the upper bound is loose). Finally, both balanced importance sampling and cross-validated importance sampling use the threshold set $\Lambda = \{0.25 \cdot \sqrt{n}, 0.5 \cdot \sqrt{n}, 1 \cdot \sqrt{n}, 1.5 \cdot \sqrt{n}, 2 \cdot \sqrt{n}\}$. These choices are based on the asymptotic considerations in \cite{ionides2008truncated} and a good default for balanced importance sampling (though see Subsection \ref{subsec:self_avoiding_walks} for a situation where this is not the case).

\begin{figure}[htpb]
	\centering
	\begin{minipage}{.48\linewidth}
	\includegraphics[width=1\textwidth]{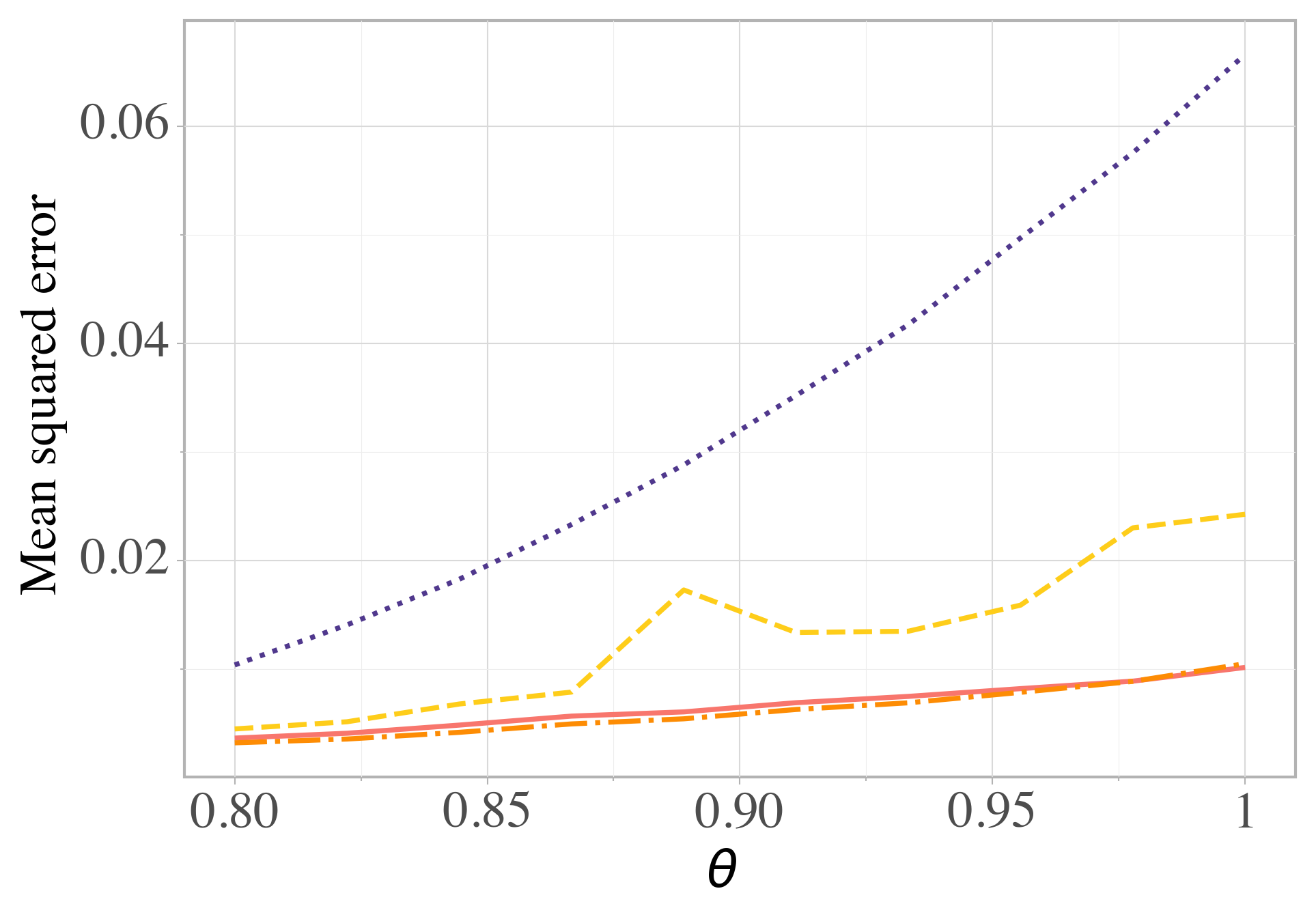}
	\includegraphics[width=1\textwidth]{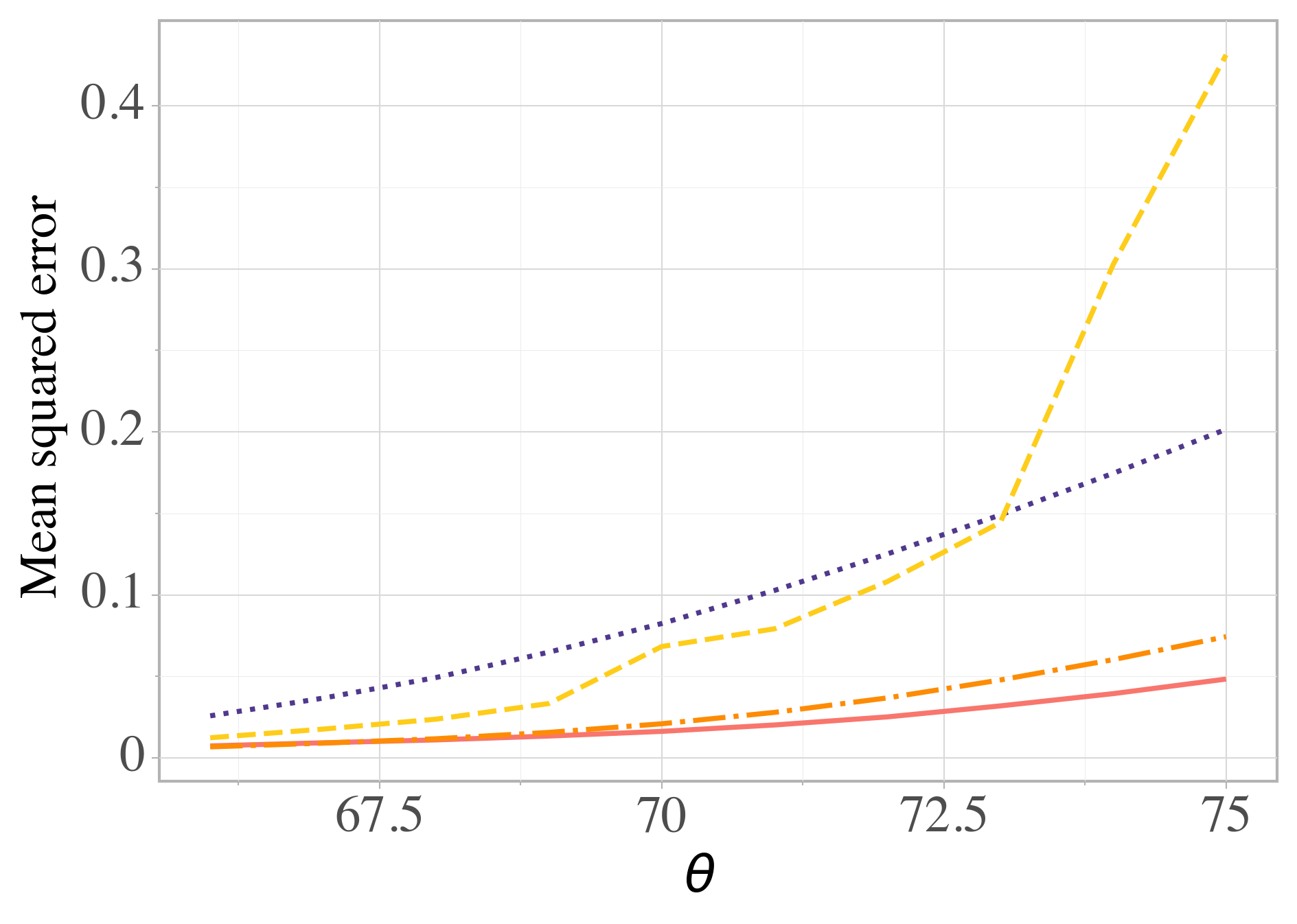}
	\includegraphics[width=1\textwidth]{exponential_MSE.png}
	\includegraphics[width=1\textwidth]{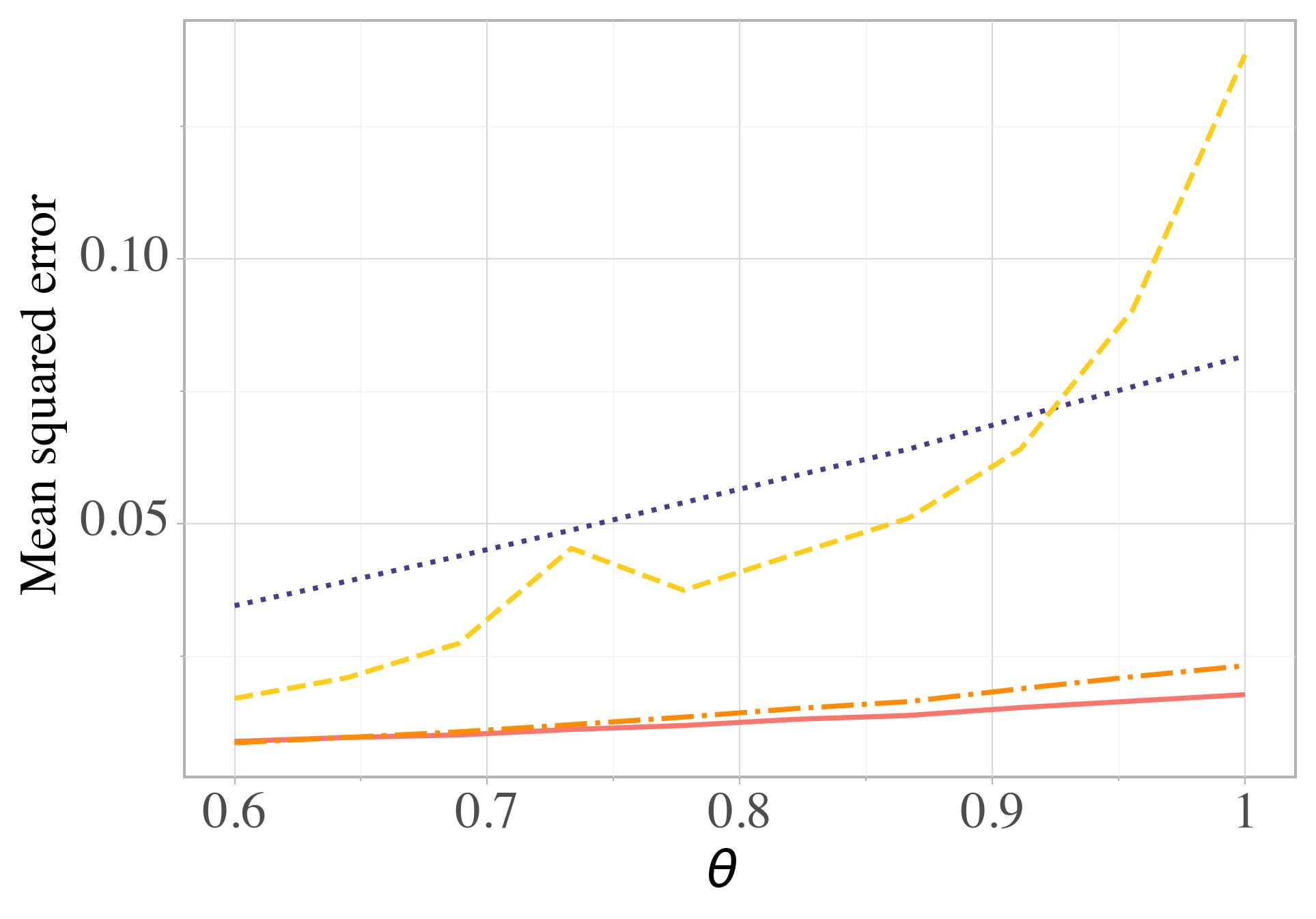}
	\end{minipage}
	\begin{minipage}{.48\linewidth}
	\includegraphics[width=1\textwidth]{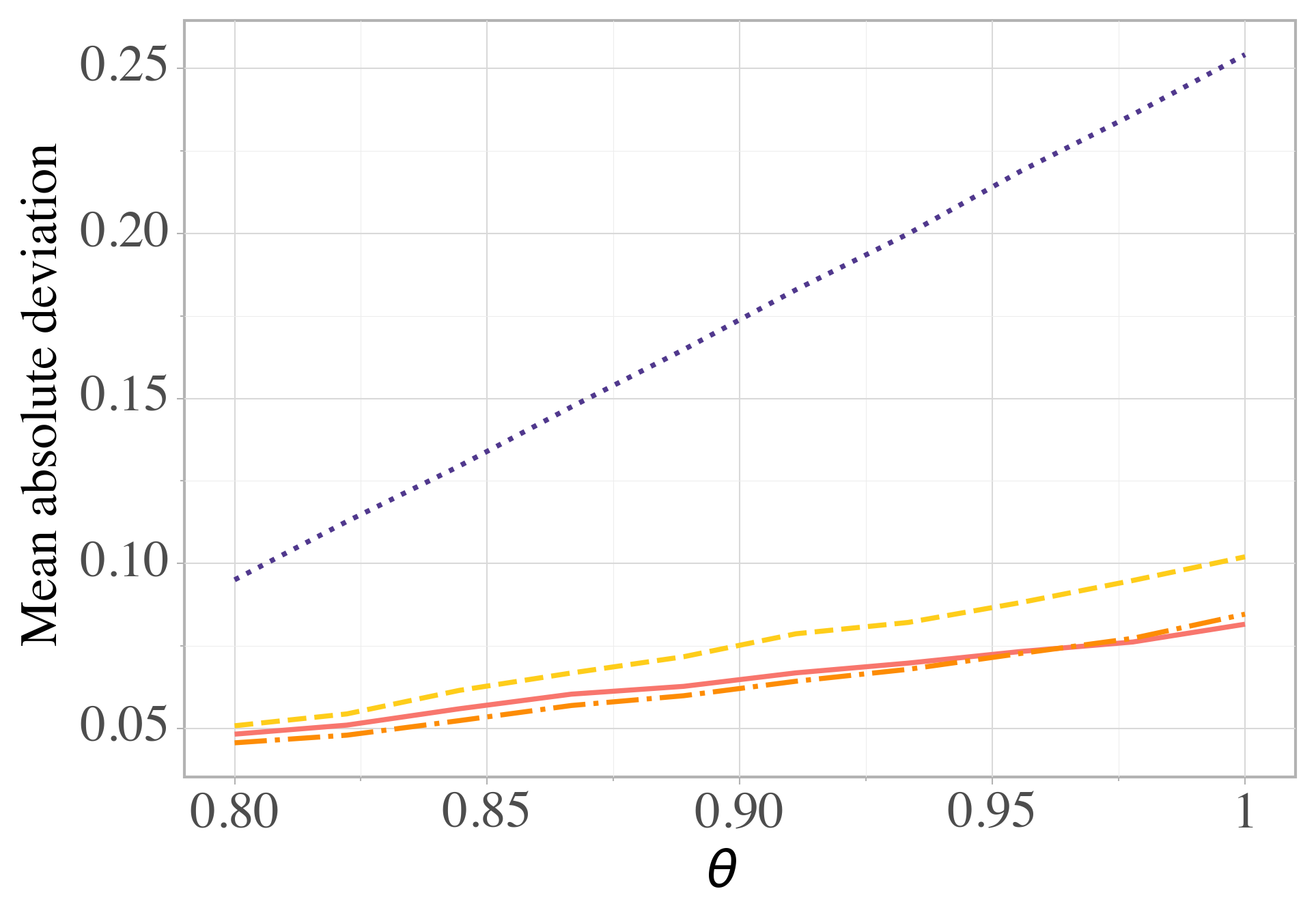}
	\includegraphics[width=1\textwidth]{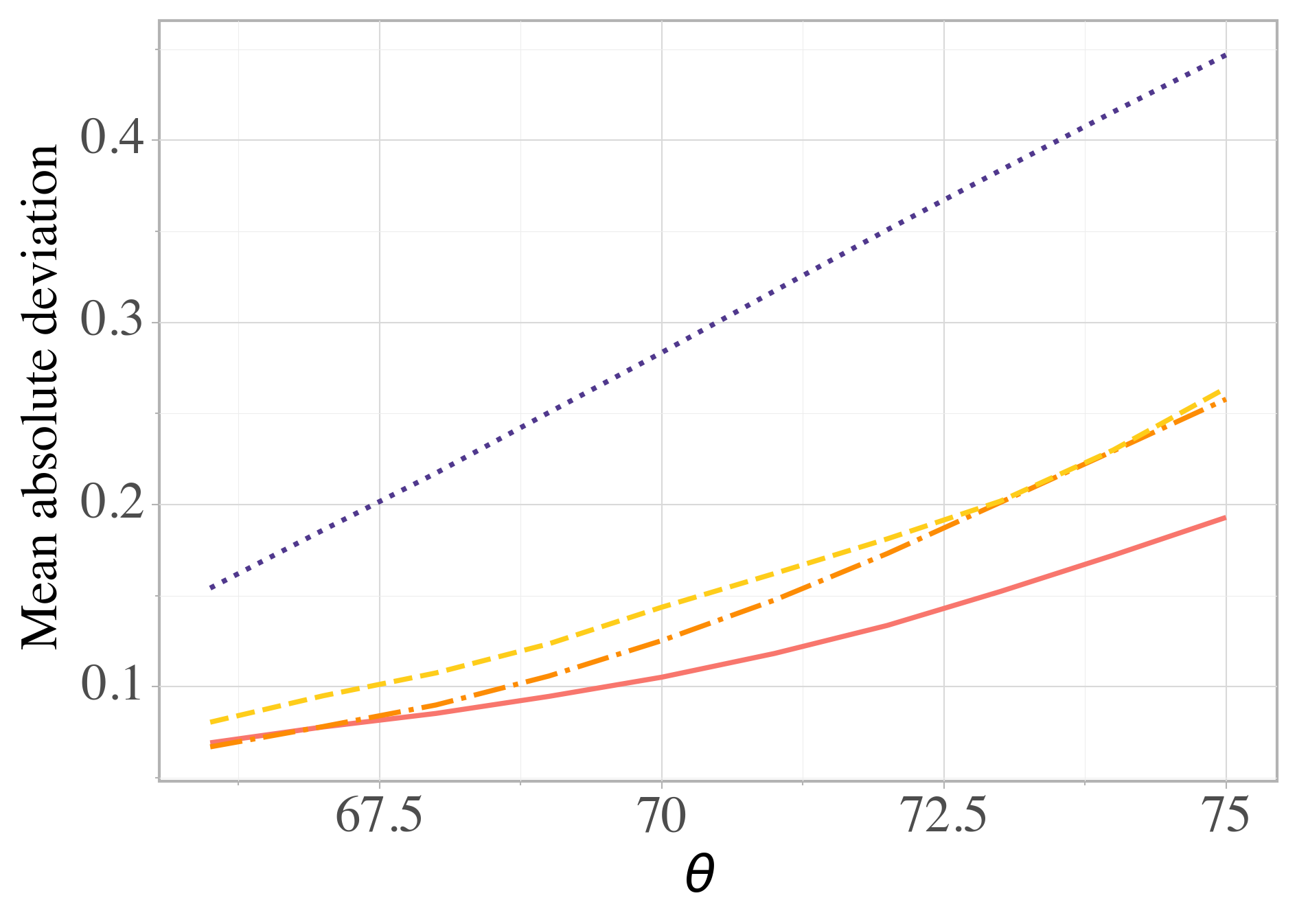}
	\includegraphics[width=1\textwidth]{exponential_MAD.png}
	\includegraphics[width=1\textwidth]{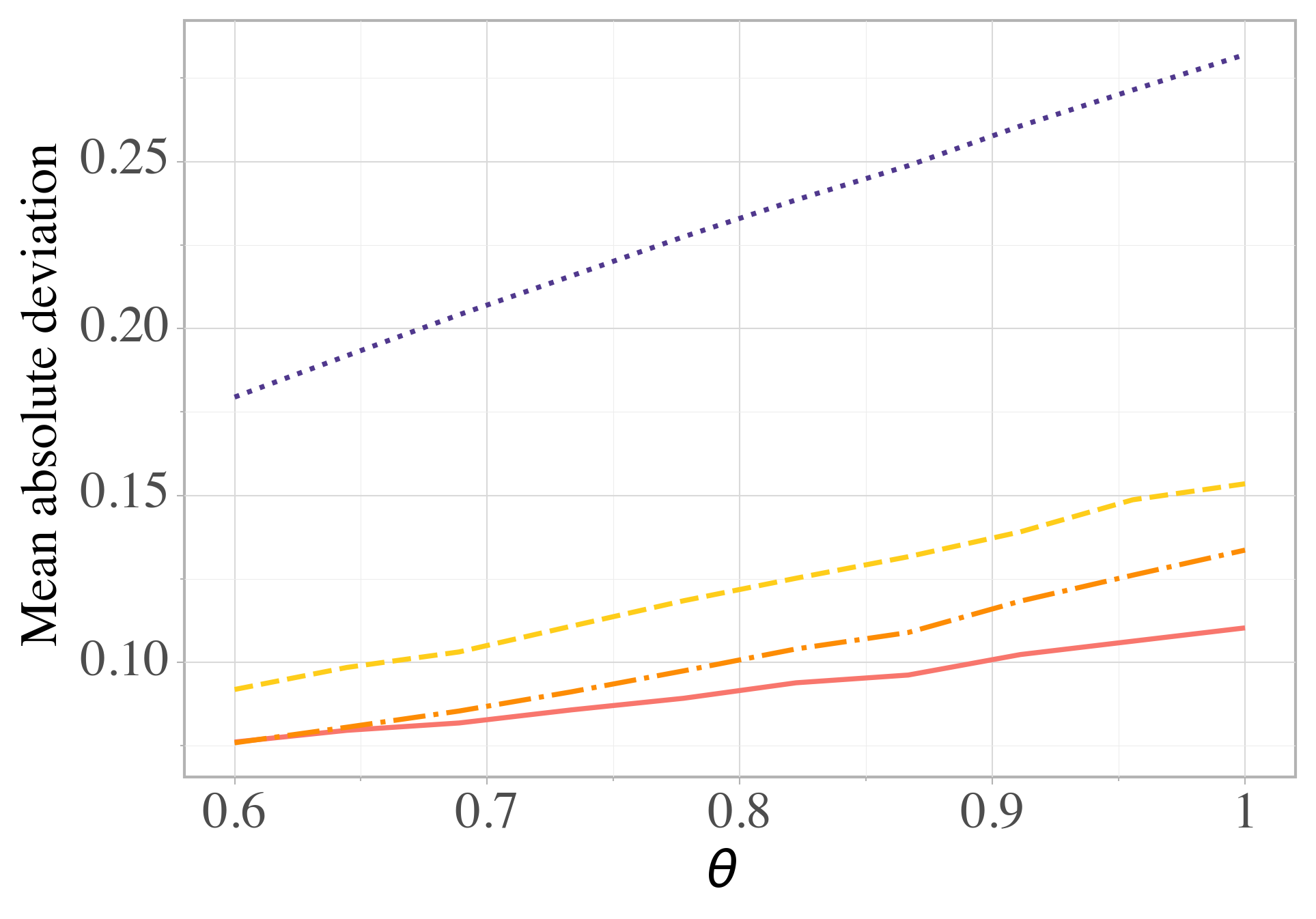}
	\end{minipage}
	\captionsetup{width=\textwidth, font=small}
	\caption{Mean squared error and mean absolute deviation errors for the beta, chi squared, exponential and normal examples, respectively. Larger values of $\theta$ indicate larger discrepancy between the proposal and target distribution. The estimators plotted are: balanced importance sampling (solid line), cross-validated importance sampling (dots), $\sqrt{n}$-winsorized importance sampling (dot-dash) and the usual importance sampling estimator (long dash).}
	\label{fig:synthetic}
\end{figure}

The results are shown in Fig.\ \ref{fig:synthetic}. In all four examples the estimation problem becomes harder as $\theta$ increases, and so the mean squared error and mean absolute deviation increase for all estimators. Balanced importance sampling is a competitive procedure, often displaying the lowest errors. In particular, Fig.\ \ref{fig:synthetic} shows the procedure is able to adequately pick the most appropriate threshold level available, which cross-validation fails to do (often overwinsorizing relative to balanced importance sampling). Also, the freedom to choose the threshold level leads to an estimator that is better than always picking $\sqrt{n}$, one of the choices available to it. Finally, balanced importance sampling often matches the performance of importance sampling in low variance settings, but becomes superior once the variance is large enough.

\subsection{Self-avoiding Walks} \label{subsec:self_avoiding_walks}

\begin{figure}[htbp]
 \centering
	\begin{minipage}{0.32\textwidth} 		\includegraphics[width=1\textwidth]{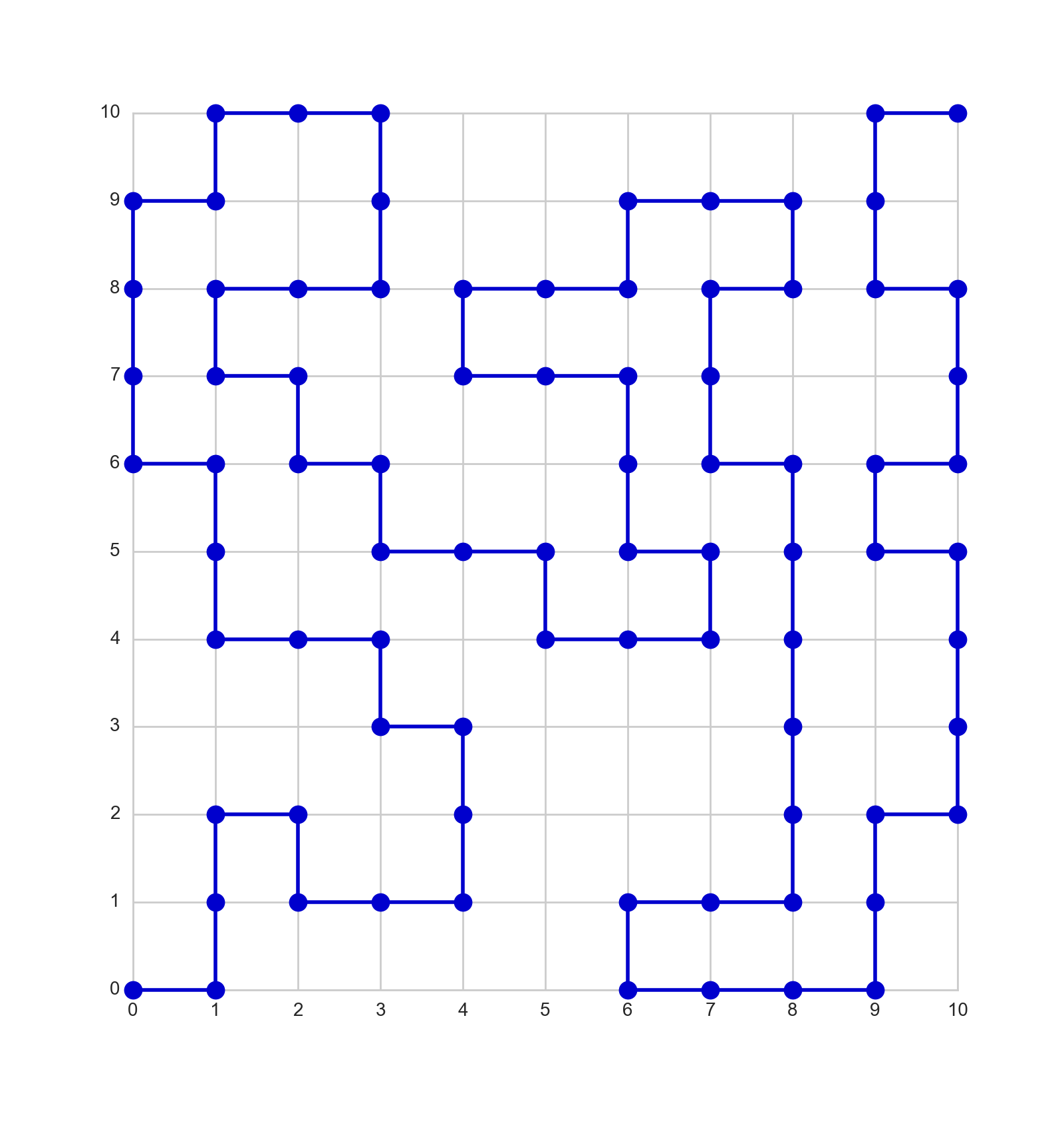}
	\end{minipage}
	\begin{minipage}{0.32\textwidth} 		\includegraphics[width=1\textwidth]{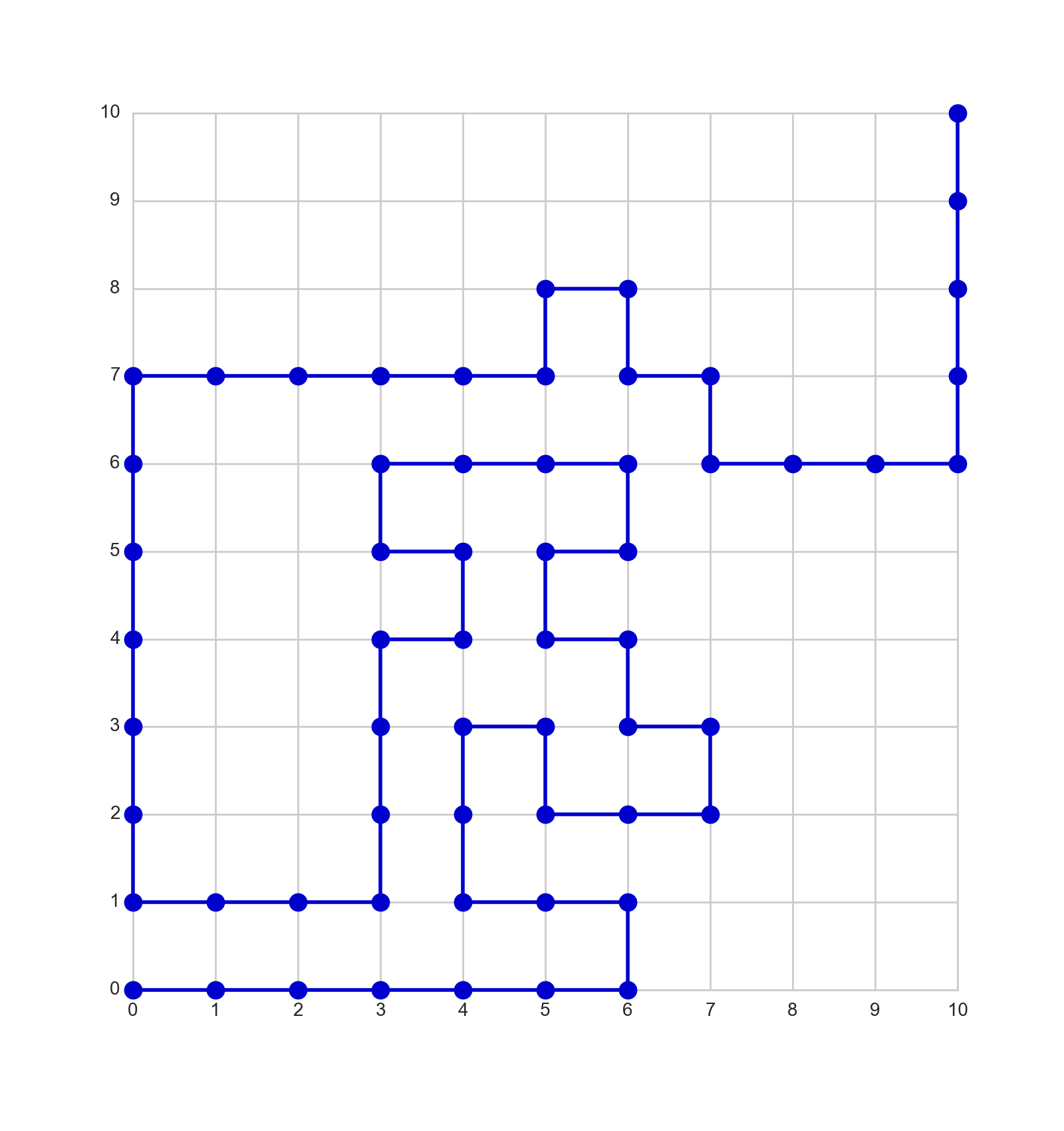}
	\end{minipage}
	\begin{minipage}{0.32\textwidth} 		\includegraphics[width=1\textwidth]{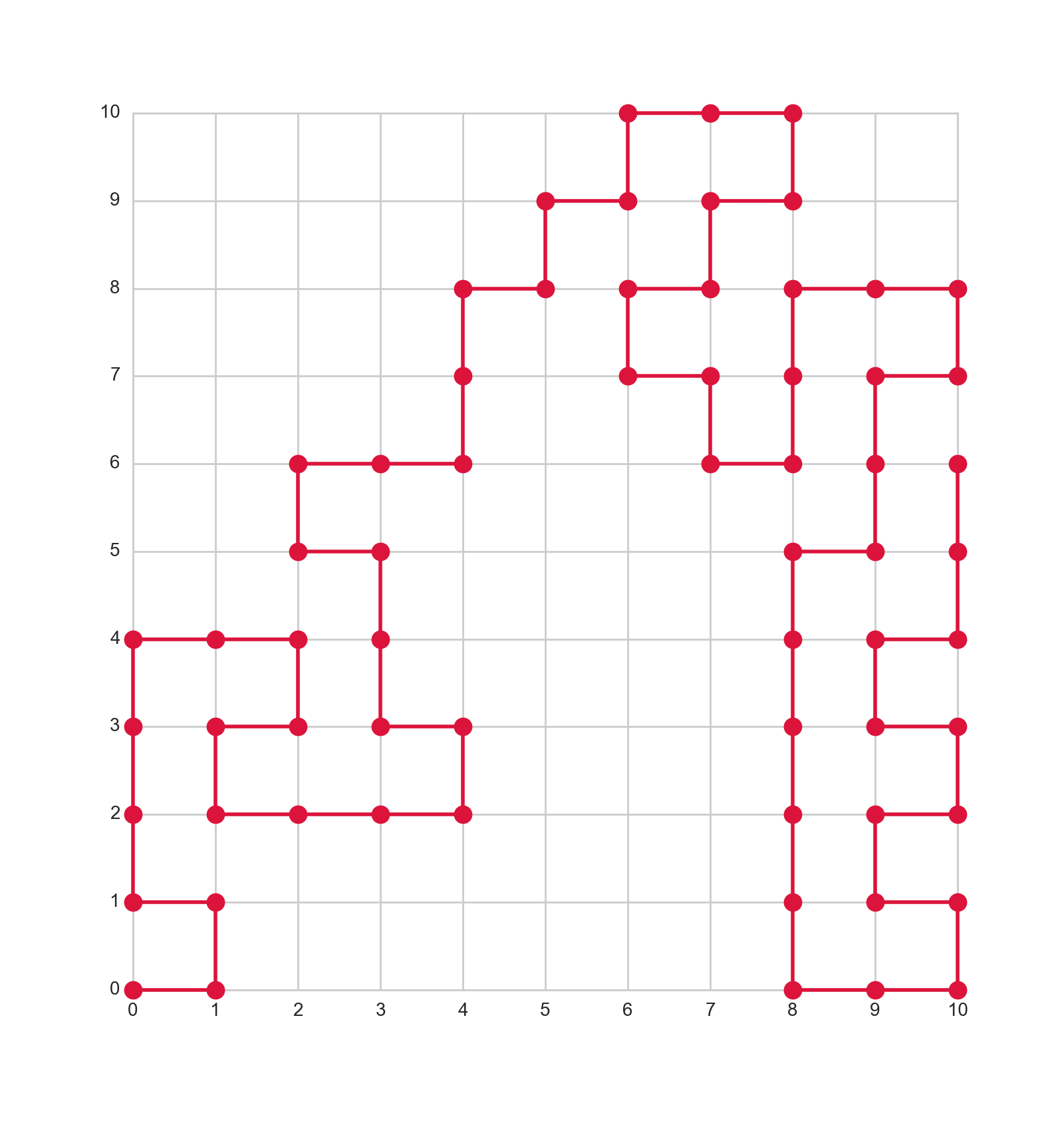}
	\end{minipage}
	\captionsetup{width=\textwidth, font=small}
	\caption{Two complete self-avoiding walks on a $10\times 10$ grid (left and center), and a walk that is forced to intersect with itself before reaching $(10, 10)$ (right).}
 \label{fig:self_avoiding_walks}
\end{figure}

As a more concrete application, consider the problem of estimating the number of self-avoiding walks on a $10 \times 10$ grid, proposed by \cite{knuth1976mathematics}. A self-avoiding walk on an $10 \times 10$ grid is a path that does not intersect itself, starting at point $(0, 0)$ and ending at point $(10, 10)$ (see Figure \ref{fig:self_avoiding_walks}). This is often taken to be a simple model for chain polymers, such as polyethylene and polyester, and are important structures in biology and chemistry. \cite{madras2013self} provides a textbook treatment of this subject.

\begin{figure}[htbp]
 \centering
	\begin{minipage}{0.325\textwidth} 		\includegraphics[width=1\textwidth]{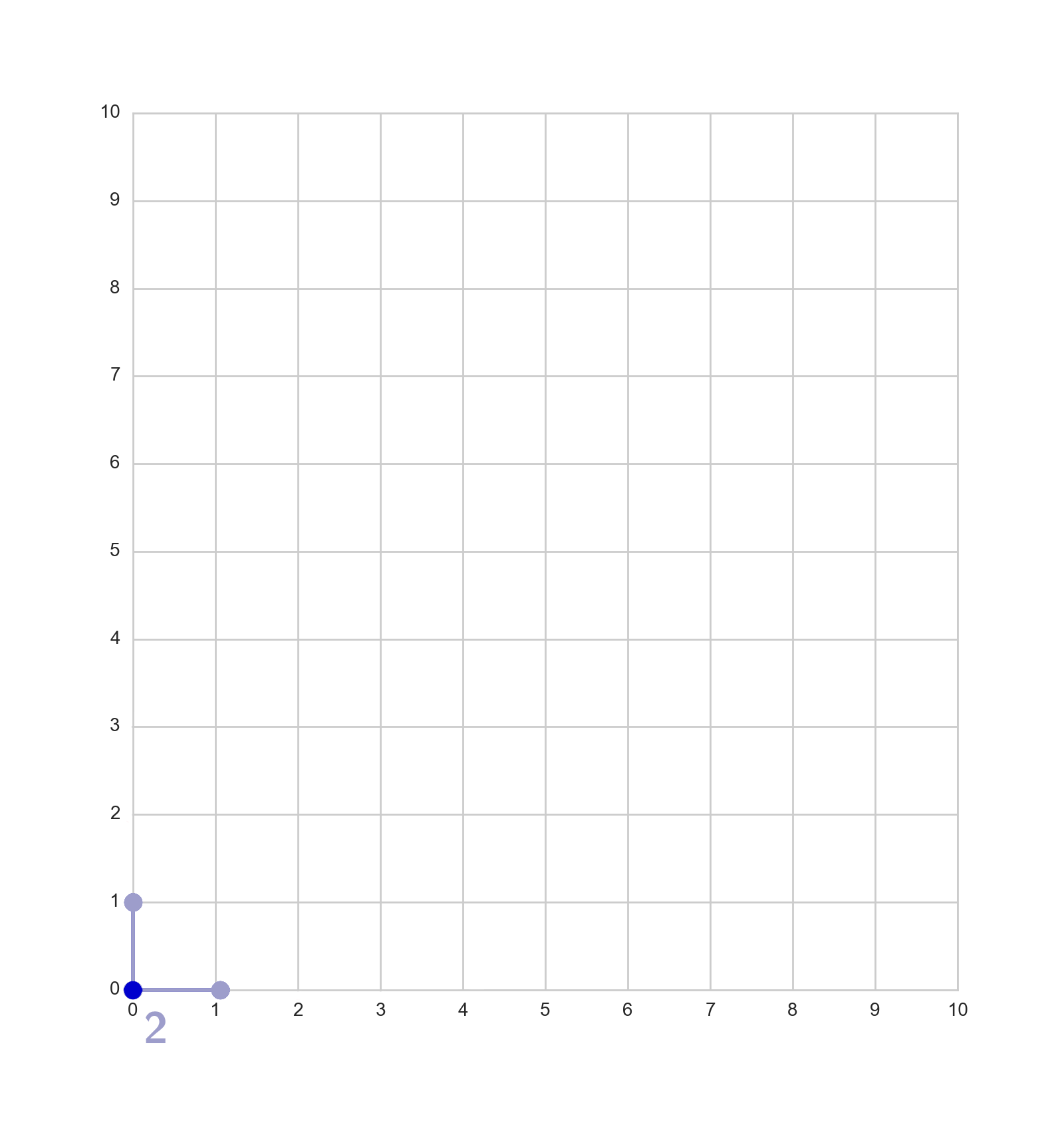}
	\end{minipage}
	\begin{minipage}{0.325\textwidth} 		\includegraphics[width=1\textwidth]{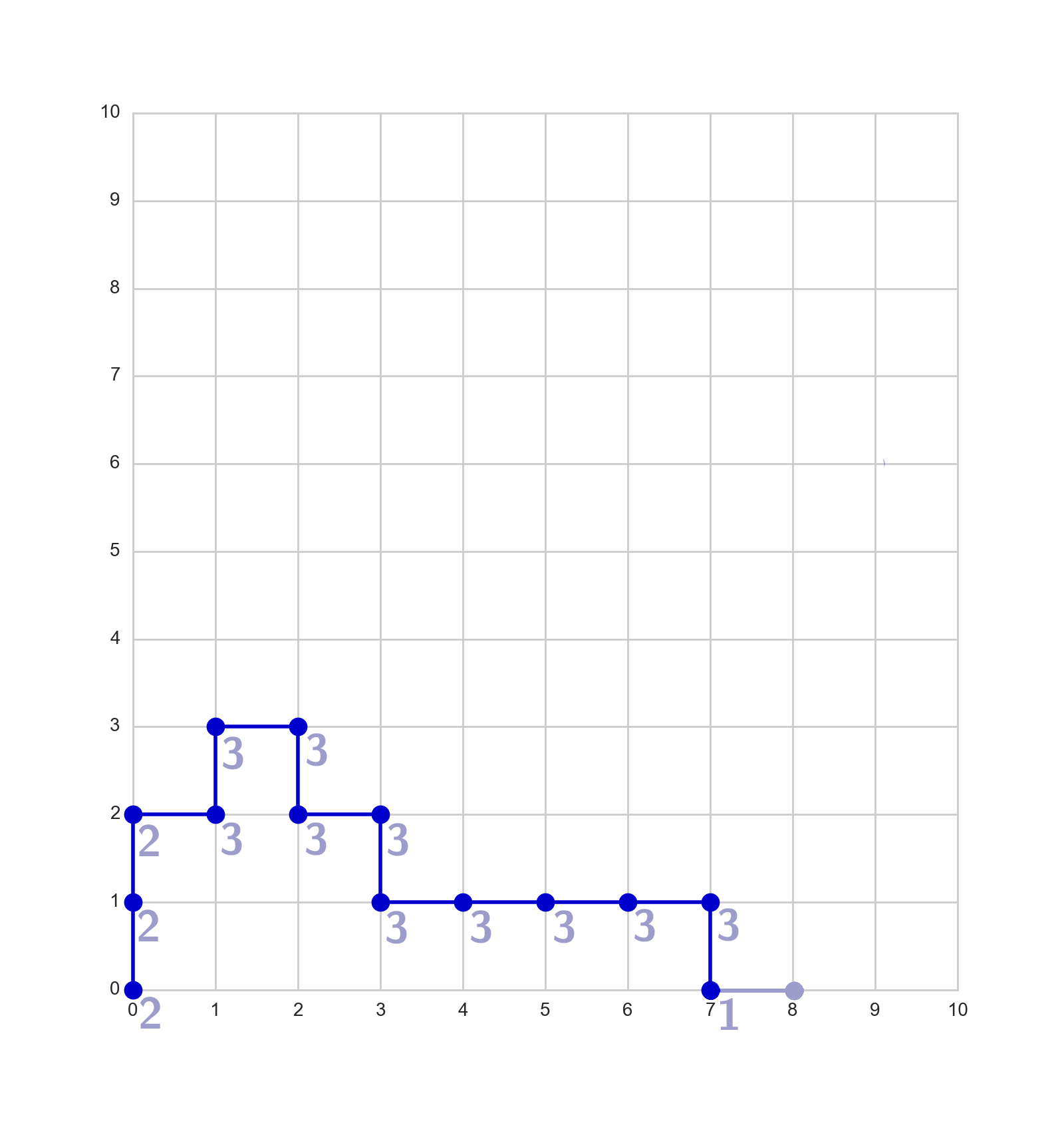}
	\end{minipage}
	\begin{minipage}{0.325\textwidth} 		\includegraphics[width=1\textwidth]{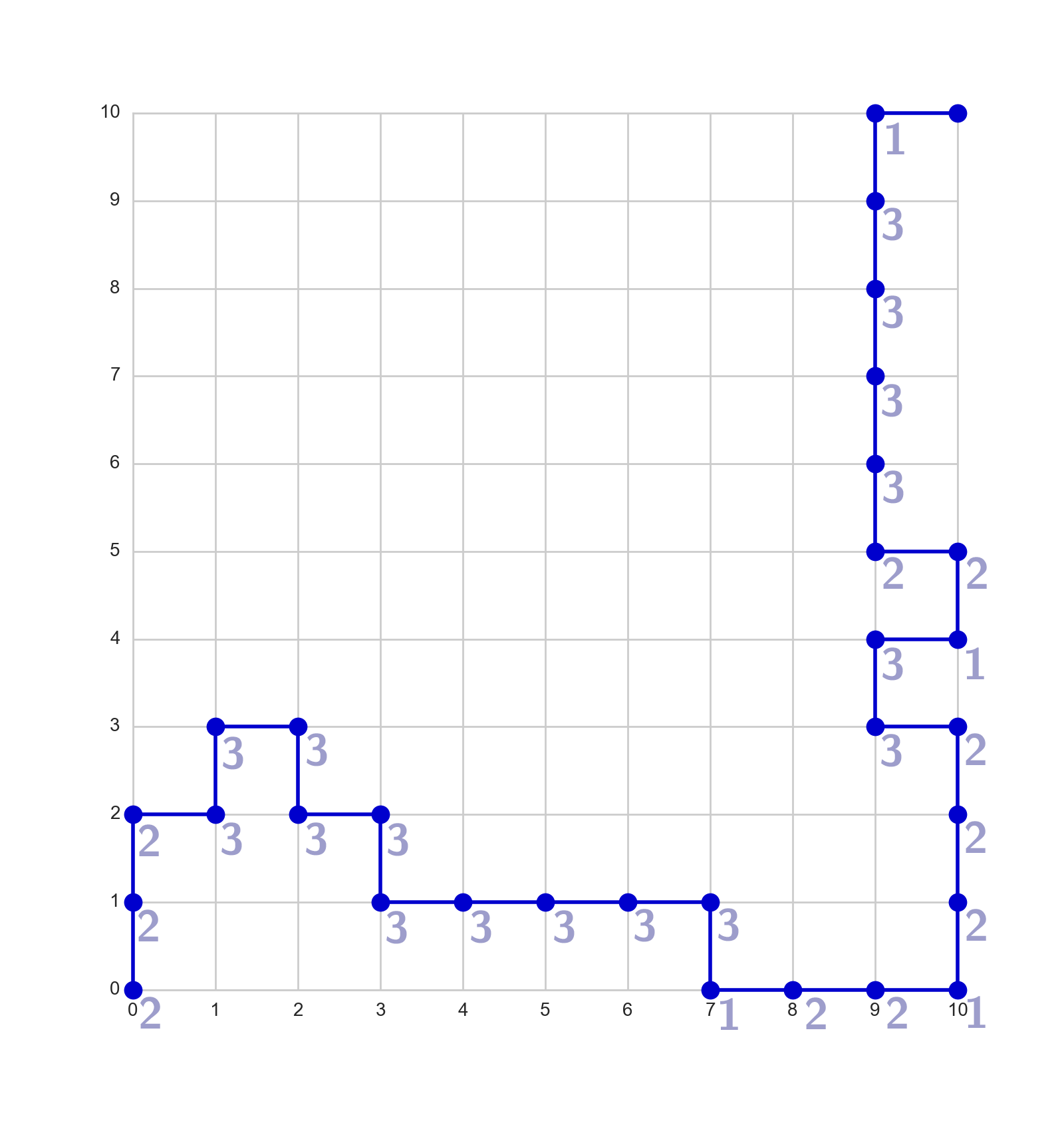}
	\end{minipage}
	\captionsetup{width=\textwidth, font=small}
 \caption{Counting the number of available neighbors $d_j$ at each point $j$ for walk $\mathbf{x}$ according to $q_3$: $d_1=2$, $d_{14}=1$, and the product is $d=\prod_{j=1}^{30}d_j= 1^{4}\times 2^{10} \times 3^{16}$; assign probability $q_3(\mathbf{x}) = 1/d = 1^{-4}\times 2^{-10} \times 3^{-16}$ to the path.}
 \label{fig:self_avoiding_walk_construction}
\end{figure}

\begin{figure}[htbp]
 \centering
	\begin{minipage}{0.32\textwidth} 		\includegraphics[width=1\textwidth]{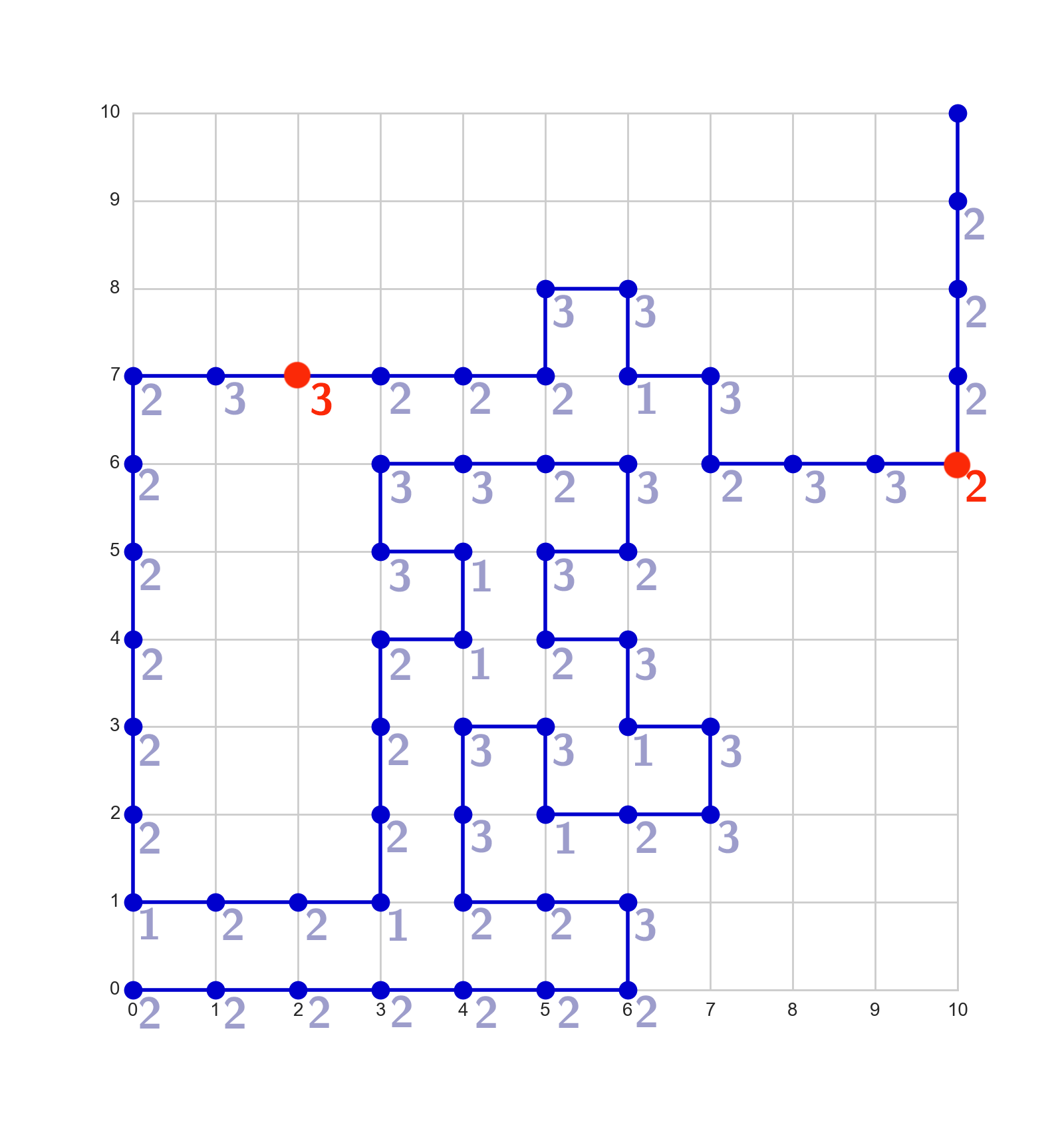}
	\end{minipage}
	\begin{minipage}{0.32\textwidth} 		\includegraphics[width=1\textwidth]{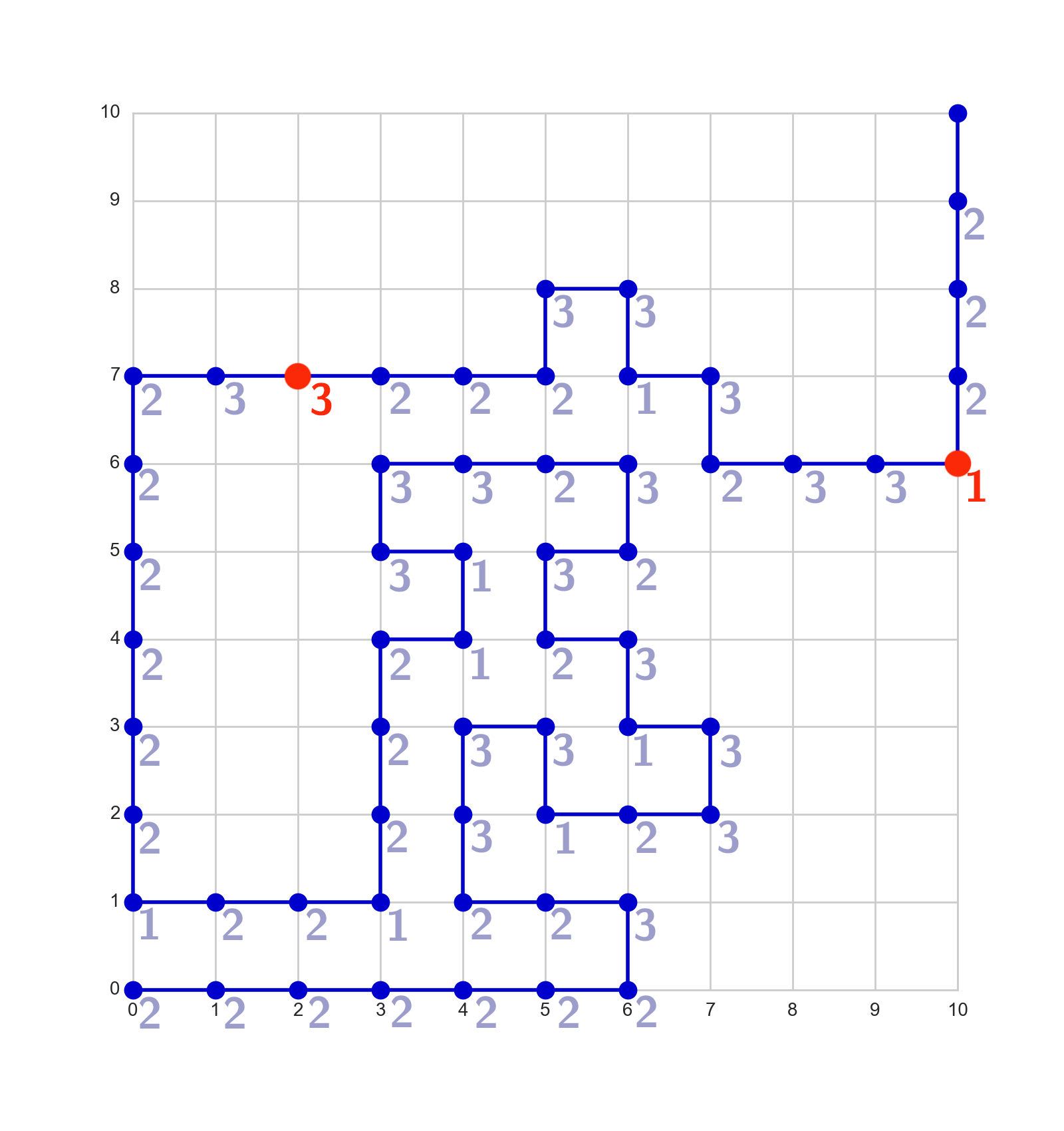}
	\end{minipage}
	\begin{minipage}{0.32\textwidth} 		\includegraphics[width=1\textwidth]{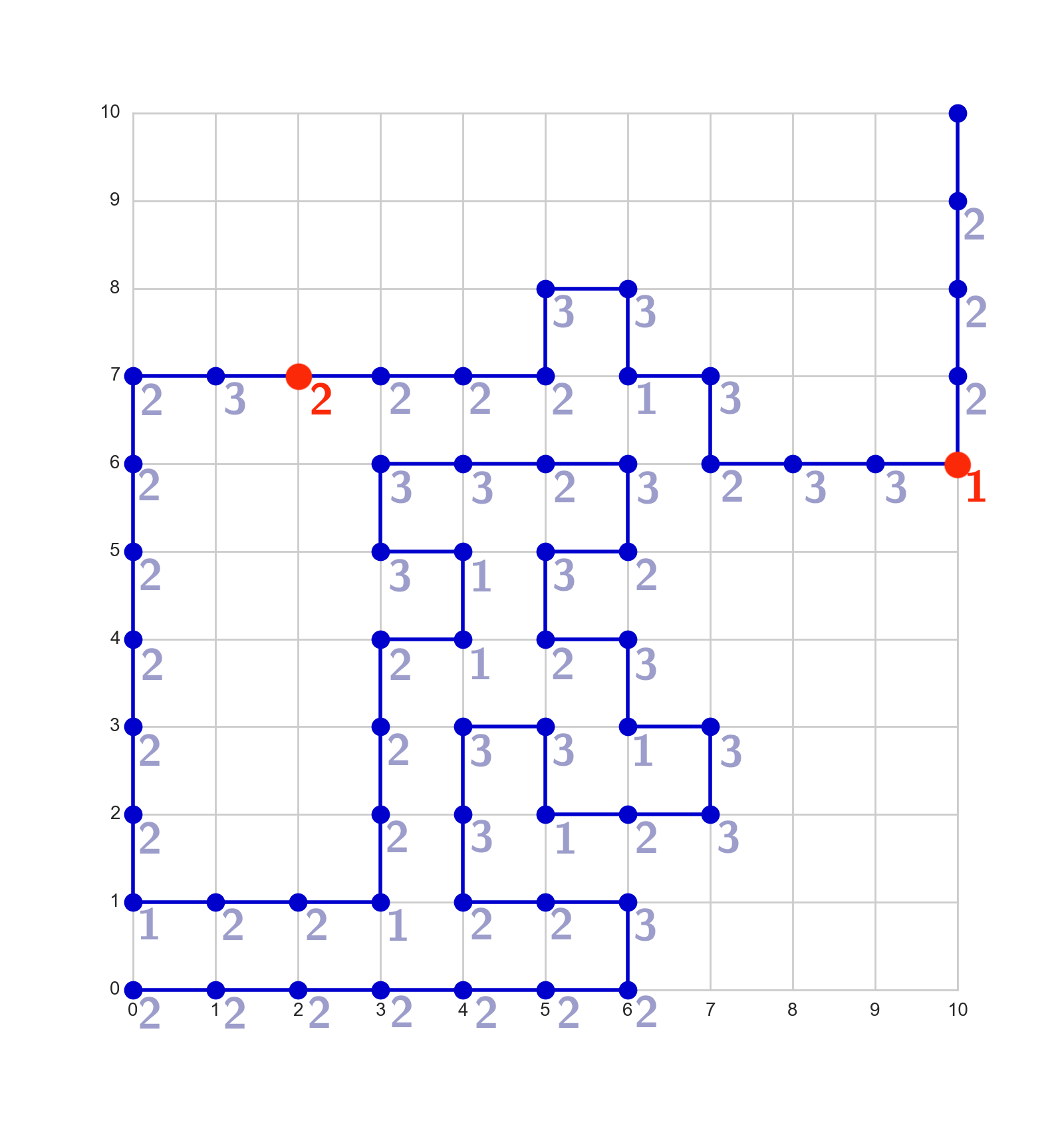}
	\end{minipage}
	\captionsetup{width=\textwidth, font=small}
 \caption{The probability of a same path $\mathbf{x}$ can be different for $q_1$ (left), $q_2$ (middle) and $q_3$ (right); the highlighted points show the differences. For $q_1$, all traps can happen; for $q_2$, there is only one available neighbor at the second red point, as going down results in a boundary trap; for $q_3$, all traps are avoided, including the interior trap that occurs if the walk goes down on the first red point. Note $q_3(\mathbf{x})<q_2(\mathbf{x})<q_1(\mathbf{x})$.}
 \label{fig:comparison_of_traps}
\end{figure}

While it is possible to find the number of $10 \times 10$ self-avoiding walks by counting every path, this becomes intractable for large grid sizes and must be estimated. Knuth proposed the following: let $q(\mathbf{x})$ be any proposal distribution on the set of walks over the $10 \times 10$ grid giving positive probability to self-avoiding walks, $p(\mathbf{x})=Z^{-1}\mathbb{I}_{[\mathbf{x} \text{ is SAW}]}$ is the uniform distribution over the self-avoiding walks, and $f(\mathbf{x})=Z$; note $Z$ counts the number of self-avoiding walks. The importance sampling estimator (\ref{eq:importance_sampling_estimator}) is:
\begin{equation*}
	\frac{1}{n} \sum_{i=1}^{n} \frac{\mathbb{I}_{[\mathbf{x}_i \text{ is SAW}]}}{q(\mathbf{x}_i)} \stackrel{n \to \infty}{\longrightarrow} \int_{\{\mathbf{x} \ : \ q(\mathbf{x}) > 0\}} \frac{\mathbb{I}_{[\mathbf{x} \text{ is SAW}]}}{q(\mathbf{x})} q(\mathbf{x}) d\mathbf{x} = \int Z p(\mathbf{x}) d\mathbf{x} = Z.
\end{equation*}
Now, consider three different proposal distributions constructed sequentially: (i) at each step $j$ in the path $\mathbf{x}$, count the number of available neighbors $d_j$ in the grid (that is, neighbors that have not been visited before and are inside the $10 \times 10$ grid), then let $q(\mathbf{x}) = \prod_{j=1}^{m_\mathbf{x}} 1/d_j$, where $m_\mathbf{x}$ is the length of walk $\mathbf{x}$; (ii) construct the distribution as in (i), but if the walk hits the any of the four boundaries do not count as available neighbors points that move closer to $(0, 0)$, as these would necessarily lead to an intersecting walk; and (iii) construct the walk as in (ii), but do not count as available neighbors any point that will force an intersecting walk. Call each of these distributions $q_1(\mathbf{x})$, $q_{2}(\mathbf{x})$ and $q_{3}(\mathbf{x})$. See Figure \ref{fig:self_avoiding_walk_construction} for an example in which $q_3(\mathbf{x})=1^{-4}\times 2^{-10} \times 3^{-16}$.
Figure \ref{fig:comparison_of_traps} highlights the differences between $q_1$, $q_2$ and $q_3$.
It is not trivial to efficiently sample from $q_3$, as it requires anticipating all moves that force an intersection; see \cite{bousquet2014importance} for details. This is representative of a large class of examples where finding better proposals can be quite hard theoretically or computationally, in which case alternatives as the one considered in this paper stand to help the most.

Using any of the three proposals $q_1(\mathbf{x})$, $q_2(\mathbf{x})$ and $q_3(\mathbf{x})$ leads to the importance sampling estimator having enormous variance, with estimates using $q_1$ worse than using $q_2$, which are worse than using $q_3$. Hence, consider applying balanced importance sampling instead.

For each of $1,000$ simulation runs, $n=10,000$ paths are generated from each of $q_1$, $q_2$ and $q_3$, and the mean squared error and mean absolute deviation from each of four procedures are considered: (i) balanced importance sampling; (ii) cross-validated importance sampling; (iii) $10^{25}$-winsorized importance sampling; and (iv) the usual importance sampling with no winsorization. The variance in this problem is large enough that winsorizing at $\sqrt{n}$ leads to a poor performance, so, since the true number of self-avoiding walks is roughly $1.56 \times 10^{24}$, the winsorization levels included are $\Lambda=\{10^{21}, 5 \cdot 10^{23}, 10^{25}, 5 \cdot 10^{26}, 10^{28}\}$, ranging from severe winsorization to virtually no capping. As in Subsection \ref{subsec:synthetic_examples}, $c=1+\sqrt{5}$ and $t=1/\sqrt{n}$.

The results are displayed in Tables \ref{table:self_avoiding_walk_mse} and \ref{table:self_avoiding_walk_mad}. The balanced procedure generally yields the best results, effectively trading-off some bias for variance, and adaptively picking better thresholds than a constant $10^{25}$. Cross-validation gives the poorest results, despite being computationally more intensive, due to its tendency to overwinsorize. The table also shows how much better the estimation under $q_3$ is for the usual importance sampling estimator. There is an improvement in mean squared error of an order of magnitude in going from $q_1$ to $q_2$, and two orders of magnitude in going from $q_2$ to $q_3$. 

\begin{table}
\def~{\hphantom{0}}
\captionsetup{width=\textwidth, font=small}
\caption{Mean squared error under three different proposals}
\begin{tabular}{lllll}
{} &            IS &   Balanced IS &         CV IS &         Fixed IS \\[5pt]
$q_1$ & 2.07 $\cdot 10^{49}$ &    1.33$\cdot 10^{48}$ & 2.46$\cdot 10^{48}$ & 2.43$\cdot 10^{48}$ \\
$q_2$ & 1.31$\cdot 10^{48}$ &    5.12$\cdot 10^{47}$ & 2.43$\cdot 10^{48}$ & 2.25$\cdot 10^{48}$  \\
$q_3$ & 3.12$\cdot 10^{46}$ &    3.07$\cdot 10^{46}$ &  2.30$\cdot 10^{48}$ & 1.36$\cdot 10^{48}$ \\
\end{tabular}
\label{table:self_avoiding_walk_mse}
\end{table}

\begin{table}
\def~{\hphantom{0}}
\captionsetup{width=\textwidth, font=small}
\caption{Mean absolute deviation under three different proposals}
\begin{tabular}{lllll}
{} &            IS &   Balanced IS &         CV IS &         Fixed IS \\[5pt]
$q_1$ & 1.77$\cdot 10^{24}$ &    1.06$\cdot 10^{24}$ & 1.57$\cdot 10^{24}$ & 1.56$\cdot 10^{24}$ \\
$q_2$ & 7.66$\cdot 10^{23}$ &    5.91$\cdot 10^{23}$ & 1.56$\cdot 10^{24}$ &  1.50$\cdot 10^{24}$ \\
$q_3$ & 1.4$\cdot 10^{23}$ &     1.4$\cdot 10^{23}$ & 1.52$\cdot 10^{24}$ & 1.17$\cdot 10^{24}$ \\
\end{tabular}
\label{table:self_avoiding_walk_mad}
\end{table}

\begin{figure}[htbp]
 \centering
	\begin{minipage}{0.32\textwidth}
		\footnotesize \centering
		$q_1$
		\includegraphics[width=1\textwidth]{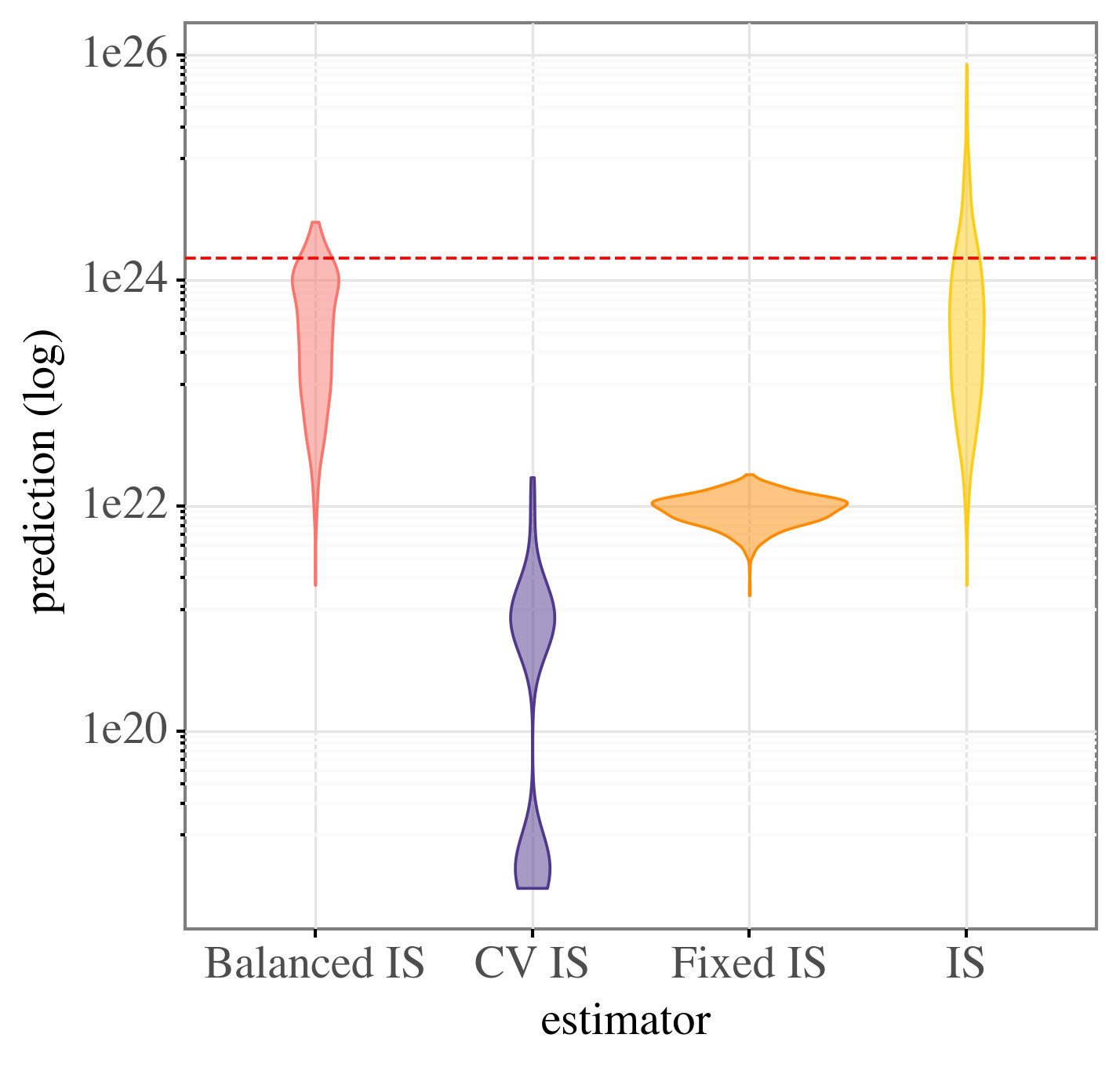}
	\end{minipage}
	\begin{minipage}{0.32\textwidth} 		
		\footnotesize \centering
		$q_2$
		\includegraphics[width=1\textwidth]{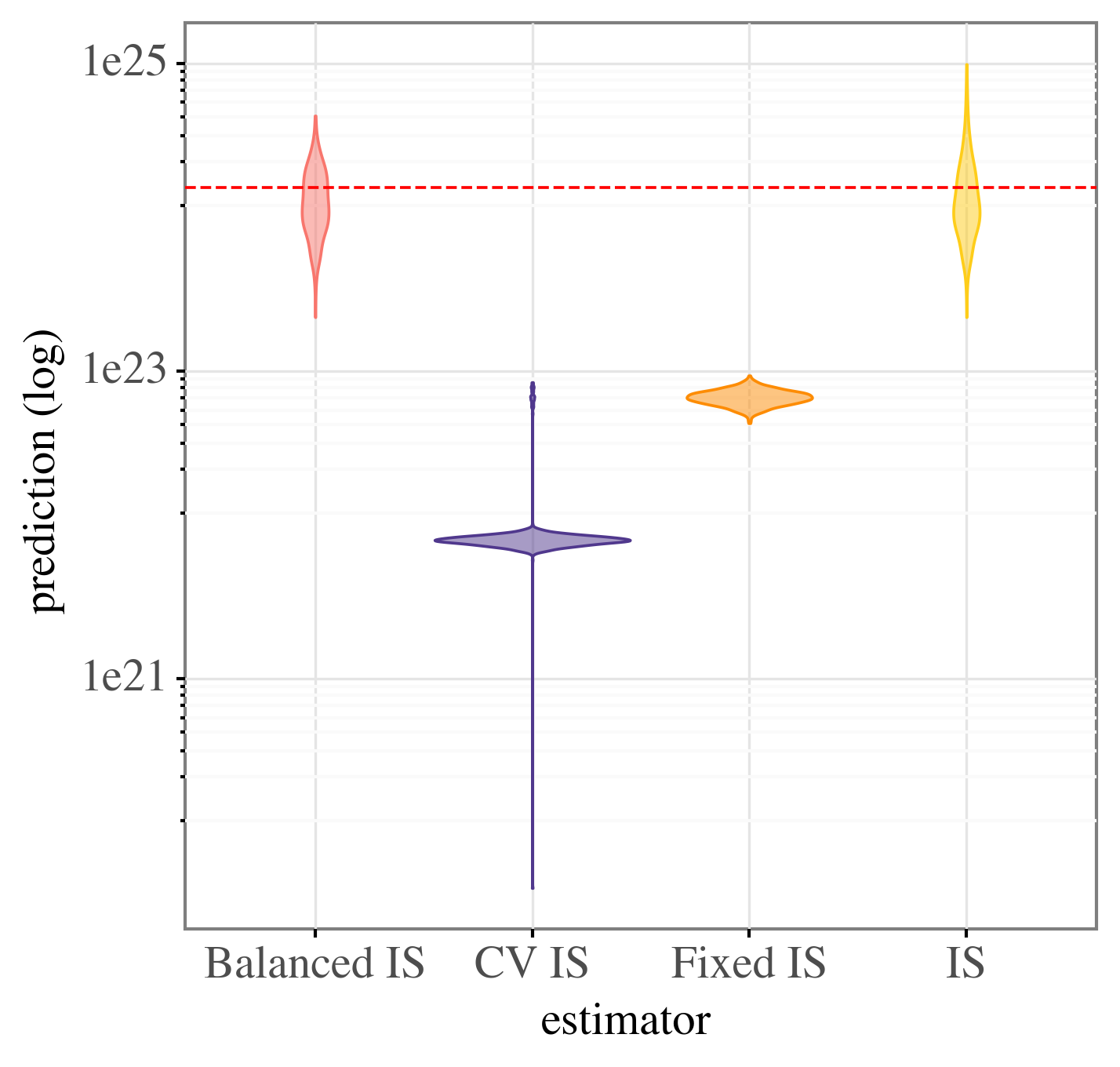}
	\end{minipage}
	\begin{minipage}{0.32\textwidth}
		\footnotesize \centering
		$q_3$
		\includegraphics[width=1\textwidth]{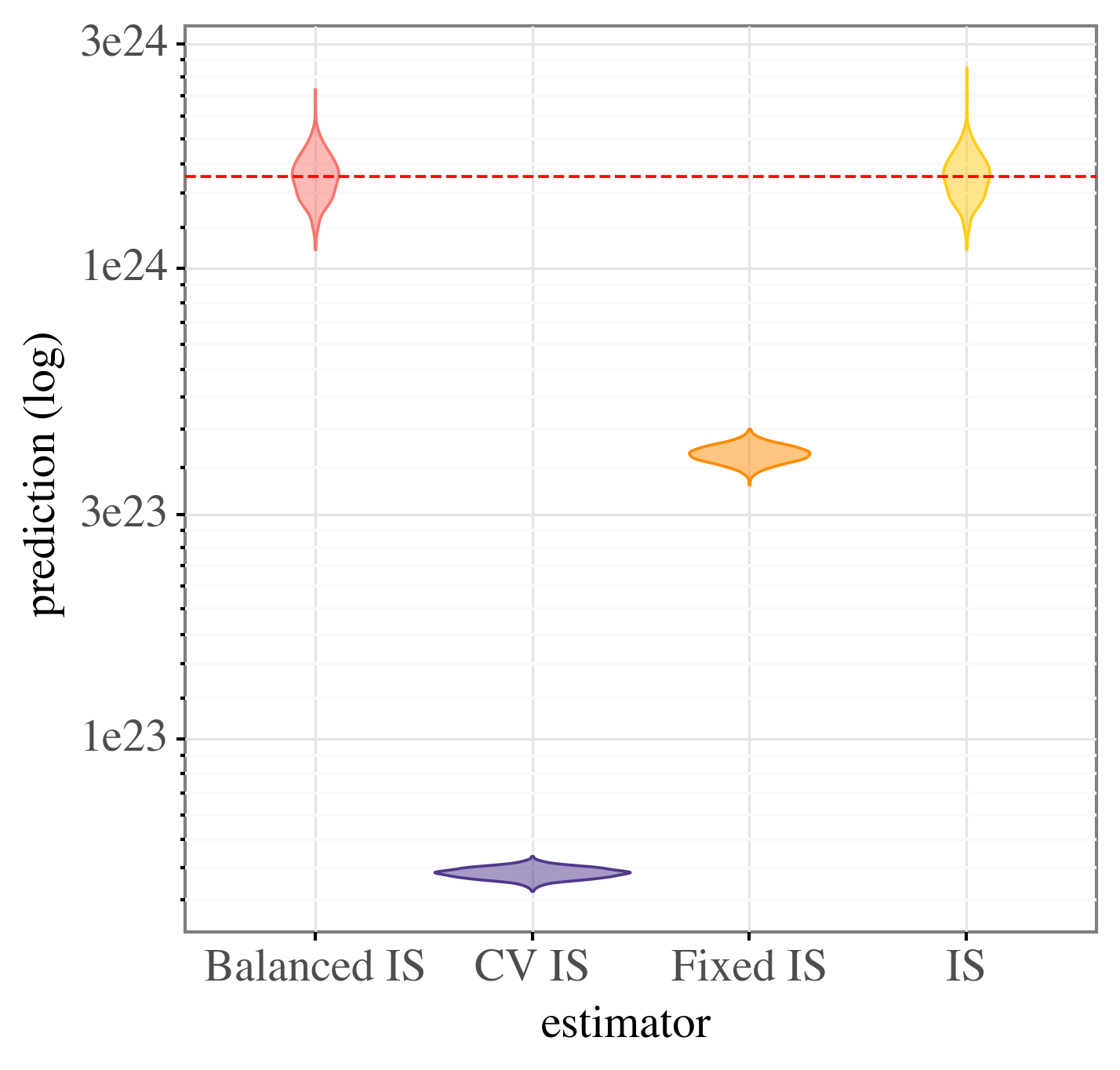}
	\end{minipage}
	\captionsetup{width=\textwidth, font=small}
	\caption{Violin plot for predictions of balanced importance sampling, winsorized importance sampling with threshold chosen via cross-validation (CV IS), winsorized importance sampling at $10^{25}$ (Fixed IS) and the importance sampling estimator with no winsorization (IS) for the three proposal distributions in the self-avoiding walk problem. The dashed horizontal line indicates the true mean. Both fixed and cross-validated estimators overwinsorize.}
	\label{fig:self_avoiding_walks-predictions}
\end{figure}

Figure \ref{fig:self_avoiding_walks-predictions} depicts the distribution for the estimates of all four methods under the proposals $q_1$, $q_2$ and $q_3$. For $q_1$, the usual importance sampling estimator includes estimates that are more than two orders of magnitude off the value to be estimated. While both cross-validated importance sampling and the importance sampling estimator winsorized at a fixed level yield estimates that are below the true mean, balanced importance sampling is able to adaptively pick threshold levels that avoid extreme realizations and efficiently trade variance for bias. 

\section{Conclusion}

While in most importance sampling settings picking a better proposal distribution is the main avenue for improving estimates, oftentimes finding such distributions can be theoretically hard or computationally challenging. In such cases, post-processing the importance sampling weights can provide an alternative and more general way to enhance the usual importance sampling estimator. This work investigated the gain obtained by winsorizing the usual estimator using a concrete version of the Balancing Principle. This yields a fast algorithm to determine the winsorization level with finite-sample optimality guarantees under minimal assumptions on the underlying data. On several examples, the procedure is shown to have good performance by matching the usual importance sampling estimator when the variance is low and offering a significant improvement when the sample variance is high.

There are many possible extensions of the method. First, theoretical results underpinning good default choices for the threshold values from a finite-sample perspective might yield a procedure that requires less user input (though the asymptotic considerations powering the choices in Subsection \ref{subsec:synthetic_examples} seem to work well). Second, using soft-thresholding instead of winsorization might lead to more conservative estimators that depend less on the winsorization level. Third, imposing further hypotheses on the problem might lead to tighter probability bounds or better procedures in particular cases. Finally, this work provides a blueprint for many other potential applications of the Balancing Principle in statistics, particularly in terms of robust mean estimation.

\section*{Acknowledgement}

The author is very grateful to Persi Diaconis and Roberto Oliveira for many helpful discussions and suggestions.

\section*{Supplementary material}
\label{SM}

Supplementary material includes counter-examples and code to reproduce the figures and table in the paper.

\subsection*{Counter-example}

While in general winsorizing the importance sampling estimator increases its bias, there are cases where this is violated, as the following simple but contrived example shows.

\begin{example}
Suppose $Y \sim \frac{1}{3} \text{Unif}[-11, -9] + \frac{1}{3} \text{Unif}[-1, 1] + \frac{1}{3} \delta_{10}$. Let $\bias(M)= |\E{Y^M}-\theta|$. In this case, $\E{Y} = \theta= 0$ and, similarly, with $M=11$ and $M=1$,
\begin{equation*}
\E{Y^{[11]}} = 0, \qquad \E{Y^{[1]}} = 0,
\end{equation*}
however, 
\begin{equation*}
\E{Y^{[10]}} = \frac{1}{3}\left(\frac{1}{2}(-10) + \frac{1}{2}(-9.5)\right) + \frac{1}{3} \cdot 0 + \frac{1}{3} \cdot 10 > 0,
\end{equation*}
so $\bias(11)=\bias(1)=0$, but $\bias(10) > 0$. Note the sufficient condition in Proposition \ref{prop:bias} is violated since $\Pr{Y\geq 10.5} < \Pr{Y \leq -10.5}$ but $\Pr{Y\geq 10}>\Pr{Y\leq -10}$.
\end{example}

\subsection*{Code}

Code implementing the proposed estimator, as well as to reproduce all the figures and tables in the paper, is available at \url{https://github.com/paulo-o/IS_adaptive_winsorization}.

\bibliographystyle{apalike}
\bibliography{paper-ref}

\end{document}